\documentclass[aps,prd,reprint,nofootinbib,superscriptaddress,amsmath,amssymb]{revtex4-2}

\usepackage[T1]{fontenc}
\usepackage[utf8]{inputenc}
\usepackage{graphicx}
\usepackage{mathrsfs}
\usepackage{amsfonts}
\usepackage{orcidlink}
\usepackage{url}
\hypersetup{hidelinks}
\allowdisplaybreaks[4]

\newcommand{\dd}{\mathrm{d}}
\newcommand{\Lned}{\mathscr{L}}

\begin{document}

\title{Two-scale magnetically charged regular black holes from nonlinear electrodynamics and a T-duality-inspired zero-point length}

\author{Ali \"Ovg\"un \orcidlink{0000-0002-9889-342X}}
\email{ali.ovgun@emu.edu.tr}
\affiliation{Physics Department, Faculty of Arts and Sciences, Eastern Mediterranean University, Famagusta, 99628 North Cyprus via Mersin 10, Turkiye.}

\author{Reggie C. Pantig \orcidlink{0000-0002-3101-8591}}
\email{rcpantig@mapua.edu.ph}
\affiliation{Physics Department, School of Foundational Studies and Education, Map\'ua University, 658 Muralla St., Intramuros, Manila 1002, Philippines.}

\author{Joel Saavedra \orcidlink{0000-0002-1430-3008}}
\email{joel.saavedra@pucv.cl}
\affiliation{Instituto de F\'{\i}sica, Pontificia Universidad Cat\'olica de Valpara\'{\i}%
so, Casilla 4950, Valpara\'{\i}so, Chile.}

\date{\today}

\begin{abstract}
We construct a two-scale, static, spherically symmetric regular black hole
in Einstein gravity sourced by magnetic nonlinear electrodynamics (NED).
The zero-point length $\ell$ regularizes the mass and charge profiles,
whereas $q$ is the asymptotic magnetic charge.  The geometry approaches
Reissner--Nordstr\"om at large radius, reduces to the neutral
zero-point-length solution for $q=0$, and coincides geometrically with the
Ay\'on-Beato--Garc\'ia solution for $\ell=|q|$.  For $q\neq0$, inverse
reconstruction gives a single-valued magnetic Lagrangian with Maxwell
asymptotics and a finite strong-field limit.  The center is regular and is
de Sitter, locally Minkowski, or anti-de Sitter according to the sign of
$2M\ell-q^2$; the weak energy condition holds globally if and only if
$3M\ell\geq2q^2$.  We derive the extremality curve, the exact heat
capacity, and homogeneous horizon-variation and Smarr identities while
retaining the Wald area entropy.  We also prove that every charged black
hole in this family has a nondegenerate extraordinary NED optical metric
throughout the domain of outer communication.  The associated capture
shadow is selected by the global minimum of the optical impact-parameter
function and generally differs from the background-geodesic shadow.
Weak-field calculations yield the periapsis, bending, time-delay, and
redshift corrections; in particular, $\ell$ first appears beyond the
standard first-post-Newtonian parameters.  Finally, for minimally coupled
test radiation in a cold transparent plasma, we obtain exact parametric
shadow relations for power-law density profiles and combine Hamiltonian
ray tracing with a Novikov--Thorne disk model.  A separate extraordinary
NED--plasma continuation is displayed only as a phenomenological
prescription because a material plasma breaks the conformal ambiguity of
the vacuum characteristic metric.
\end{abstract}

\maketitle

\section{Introduction}
\label{sec:introduction}

The occurrence of curvature singularities in classical black-hole
solutions is generally interpreted as an indication that the
Einstein--Maxwell description ceases to be reliable in the strong-field
ultraviolet regime. This has motivated extensive efforts to construct
regular black-hole geometries in which the metric and curvature
invariants remain finite throughout the spacetime. Early examples include
the phenomenological Bardeen model, the Dymnikova geometry, and the
Hayward spacetime \cite{Bardeen1968,Dymnikova1992,Hayward2006}. A major
development was the demonstration that regular charged geometries can
arise within Einstein gravity coupled to nonlinear electrodynamics
(NED) \cite{ABG1998,ABG1999,Ayon-Beato:2000mjt,Bronnikov2001}.
Subsequent studies have clarified both the usefulness and the limitations
of NED as a source of regular black holes
\cite{BronnikovReview2022,Bokulic:2024reverse}.

The magnetic sector is particularly important. For a theory with a
Lagrangian depending on the invariant
$F=F_{\mu\nu}F^{\mu\nu}$ and possessing the Maxwell weak-field limit,
globally regular solutions are naturally associated with magnetic rather
than purely electric charge \cite{Bronnikov2001,BronnikovReview2022}.
Conversely, the reconstruction of a matter Lagrangian from a prescribed
regular metric is an inverse problem and does not automatically guarantee
that the resulting NED is a universal microscopic theory. In many
regular-black-hole models, parameters that are normally regarded as
integration constants also occur explicitly in the reconstructed
Lagrangian \cite{Bokulic:2024reverse,Guo:2024thermo}. This feature must be
taken into account when interpreting the matter sector and its
thermodynamics.

An independent route to regular short-distance geometries is provided by
the zero-point-length prescription associated with path-integral duality.
Padmanabhan showed that imposing invariance of the worldline amplitude
under the transformation
${\cal R}\rightarrow L_0^2/{\cal R}$ effectively replaces the squared
geodesic distance by a quantity containing a nonzero minimum length
\cite{Padmanabhan1997,Padmanabhan1998}. This construction is closely
related to string T-duality and provides a phenomenological description
of ultraviolet suppression in propagators
\cite{NicoliniReview2022}. Applied to the Newtonian potential, it produces
a regular neutral black-hole metric that is formally of Bardeen type,
with the magnetic charge replaced by the zero-point length
\cite{Nicolini2019}.

The zero-point-length prescription has also been extended to the
electromagnetic sector. A nonlocal, gauge-invariant deformation of
electrodynamics regularizes the Coulomb potential and suppresses the
electric field near the origin \cite{Gaete2022}. Charged and rotating
black-hole geometries, wormholes and other various type of black-hole geometries, derived from the same T-duality-motivated
propagator were subsequently presented in
~\cite{Gaete:2022ukm,Lutfuoglu:2026etg,Orzuev:2026szf,Javed:2024ocf,Javed:2024wbc,Eghbali:2023sak}. In the static sector, that construction is
closely related to the Ay\'on-Beato--Garc\'ia geometry when the
zero-point-length parameter is identified with the electromagnetic
charge. These results establish the existence of charged
T-duality-inspired regular black holes but leave open the complementary
question of whether the charge and the ultraviolet length can be retained
as independent parameters within a local magnetic NED representation.

There is also a direct observational motivation for studying the NED
sector rather than only the background metric. In nonlinear
electrodynamics, high-frequency electromagnetic perturbations generally
propagate along null geodesics of an effective optical geometry, not
along null geodesics of the spacetime metric
\cite{Novello2000,Toshmatov2018}. This distinction has been shown to
produce appreciable modifications of photon orbits, shadows, and lensing
in regular NED black holes
\cite{StuchlikSchee2019,Toshmatov2021,dePaula:2023nedshadow}. The
difference is physically important in the context of horizon-scale
imaging: the Event Horizon Telescope observations of M87$^*$ and
Sgr~A$^*$ have transformed the black-hole shadow into an observable
strong-field diagnostic
\cite{EventHorizonTelescope:2019M87,EventHorizonTelescope:2022SgrA,Battista:2026nsx}.
For general analytical treatments of black-hole shadows and their
relation to unstable photon orbits, see
~\cite{Perlick:2021aok,Pantig:2025shadow}.

In an astrophysical environment, however, the observed image is not
determined by vacuum photon propagation alone. A dilute electron plasma
introduces a local plasma frequency and a frequency-dependent refractive
index, so that the circular photon orbit, critical impact parameter, and
apparent shadow become chromatic
\cite{Perlick:2015vta,Kobialko:2025plasma}. At sufficiently low
frequencies, dispersive reflection can prevent rays from reaching the
strong-field region and can drive the critical capture impact parameter
to zero. The disappearance of the central capture shadow does not imply
that the surrounding accretion flow becomes invisible, since direct and
plasma-reflected disk rays may still reach the observer. A consistent
comparison with observations therefore requires the shadow calculation
to be supplemented by a model of the emitting matter and by relativistic
radiative transfer.

Thin accretion disks provide a framework for this purpose.
In the Novikov--Thorne approximation, the radial energy flux and
temperature are determined by the circular timelike geodesics of the
background geometry, while the observed intensity additionally depends
on gravitational and Doppler redshifts, plasma dispersion, and the
mapping between the disk and the observer's screen
\cite{Page:1974he,Bambi:2017khi}. In a transparent dispersive medium,
the invariant photon distribution function can be transported along the
Hamiltonian rays, allowing frequency-dependent intensity maps and
spectra to be obtained by backward ray tracing
\cite{Lindquist:1966igj,Kichenassamy:1985zz,Kobialko:2025plasma}.
Such observables contain information that is absent from the vacuum
shadow radius alone, because they depend on the disk dynamics, the
density profile of the plasma, and the observing frequency.

Combining a material plasma with nonlinear-electrodynamic photon
propagation requires an additional qualification. In vacuum, the
extraordinary NED characteristic equation determines an effective
optical metric only up to a conformal transformation, since conformally
related metrics possess the same null trajectories. The plasma-frequency
term is dimensionful and is not conformally invariant. Consequently, a
unique dispersion relation combining the extraordinary NED cone with a
material plasma cannot be inferred from the vacuum effective metric
alone; it requires a microscopic constitutive model incorporating both
the nonlinear vacuum response and the dielectric response of the
medium. We therefore distinguish two levels of analysis. The thin-disk
images and emission spectra are calculated for minimally coupled test
electromagnetic radiation propagating through an external plasma on the
background spacetime. In parallel, we introduce a minimal
phenomenological continuation of the extraordinary NED cone to examine
its possible influence on the frequency-dependent shadow. The latter is
used as an optical diagnostic and is not presented as a unique
microscopic NED--plasma theory.

The thermodynamics of regular NED black holes likewise requires special
care. When the NED Lagrangian contains parameters such as the mass or
charge, direct use of the standard first law can produce an entropy that
does not agree with the Bekenstein--Hawking area formula. Covariant and
extended-phase-space treatments show that these parameters must be varied
as couplings of the matter theory, leading to additional work terms and
a modified Smarr relation \cite{Breton2005,Guo:2024thermo}. Recent
approaches based on nonextremality and scaling further illustrate how
thermodynamic information can be reconstructed from asymptotic charges,
horizon data, and homogeneity properties
\cite{Yang:2025rud}. In the present model, the independent dimensionful
scale $\ell$ and the explicit mass dependence of the reconstructed NED
sector require a corresponding generalization.

Motivated by these considerations, we derive the line element
\begin{equation}
 \dd s^2
 =
 -f(r)\dd t^2
 +\frac{\dd r^2}{f(r)}
 +r^2\dd\Omega^2,
 \label{eq:metric}
\end{equation}
where
\begin{equation}
 f(r)
 =
 1-\frac{2Mr^2}{(r^2+\ell^2)^{3/2}}
 +\frac{q^2r^2}{(r^2+\ell^2)^2}.
 \label{eq:newmetric}
\end{equation}
The parameters $q$ and $\ell$ remain independent at the level of the
metric. The first term is inherited from the T-duality-inspired neutral
mass profile, while the second is chosen so that the geometry is regular
at the center and approaches the Maxwell charge contribution at large
radius. The metric reduces to the neutral zero-point-length geometry for
$q=0$, approaches Reissner--Nordstr\"om for $\ell\to0$ at fixed
$r\neq0$, and reproduces the Ay\'on-Beato--Garc\'ia metric function when
$\ell=|q|$. These are geometric limits; they do not imply that the
underlying electromagnetic theories are identical.
Throughout, we take $M>0$, $\ell>0$, and $q\in\mathbb{R}$; all background
quantities are invariant under $q\mapsto-q$.

For $q\neq0$, we reconstruct the magnetic NED Lagrangian supporting
Eq.~\eqref{eq:newmetric}. We establish its Maxwell weak-field limit,
analyze its strong-field behavior, and derive the exact condition for
global satisfaction of the weak energy condition. We then determine the
horizon structure and the extremal curve. The thermodynamic analysis
retains the Wald area entropy and treats the zero-point length as an
additional scaling variable, leading to generalized first-law and
Smarr-type relations, an exact heat capacity, and local-stability
criteria.

In the vacuum optical sector, we derive the effective NED geometry and
obtain the corresponding light-ring and shadow equations. We explicitly
distinguish the extraordinary NED shadow from the geometric shadow
defined by null geodesics of the background metric and calculate their
leading difference about the Reissner--Nordstr\"om limit. We subsequently
embed the compact object in a cold, nonmagnetized, transparent plasma.
For inverse power-law plasma profiles, we derive the
frequency-dependent circular-ray and shadow relations and identify the
frequency at which the central capture shadow disappears. We combine
the dispersive Hamiltonian propagation with a Novikov--Thorne disk model
and invariant radiative transfer to compute frequency-dependent
intensity maps and emission spectra by backward ray tracing. This
construction separates the modification produced by the regular
background geometry, the extraordinary NED optical response, and the
environmental dispersion generated by the plasma.

It is important to note the scope of the paper. Since the
reconstructed function $\Lned(F)$ depends explicitly on $M$, $q$, and
$\ell$, the model should be viewed as a parameter-dependent effective
NED representation of the geometry rather than as a family of states
generated by one universal microscopic electromagnetic Lagrangian. In
addition, the magnetic reconstruction is defined only for $q\neq0$.
The limit $q\to0$ is smooth at the level of the metric but is not a
regular limit of the magnetic inverse-NED formulas.

The paper is organized as follows. In
Sec.~\ref{sec:einstein-ned} we introduce the Einstein--NED equations and
the inverse magnetic reconstruction. Section~\ref{sec:solution} presents
the geometry, its regularity properties, the energy conditions, and the
explicit NED Lagrangian. The horizon structure and extremality are
studied in Sec.~\ref{sec:horizons}. In Sec.~\ref{sec:thermo} we analyze
the temperature, entropy, generalized thermodynamic relations, heat
capacity, and stability.  Section~\ref{sec:physical_parameter_map}
establishes the physical parameter domain and proves exterior optical
admissibility.  Weak-field classical tests are derived in
Sec.~\ref{sec:classical_tests}.  Section~\ref{sec:shadow} then develops
the vacuum effective optical geometry and compares the extraordinary NED
and background-geodesic shadows.  In Sec.~\ref{sec:plasma_disk} we investigate frequency-dependent
propagation through a cold plasma and calculate Novikov--Thorne disk
images and emission spectra. We summarize the results in
Sec.~\ref{sec:discussion}. Throughout the paper we use
$G=c=\hbar=k_B=1$ and the metric signature $(-,+,+,+)$.

\section{Einstein-NED system}
\label{sec:einstein-ned}

We start from the action
\begin{equation}
 I=\frac{1}{16\pi}\int \dd^4x\sqrt{-g}\left[R-\Lned(F)\right],
 \label{eq:action}
\end{equation}
where
\begin{equation}
 F=F_{\mu\nu}F^{\mu\nu},
 \qquad
 F_{\mu\nu}=\partial_\mu A_\nu-\partial_\nu A_\mu .
\end{equation}
The field equations are
\begin{align}
 G_{\mu}^{\ \nu}&=8\pi T_{\mu}^{\ \nu},
 \label{eq:einstein}\\
 \nabla_\mu\left(\Lned_F F^{\mu\nu}\right)&=0,
 \label{eq:nedmaxwell}
\end{align}
where $\Lned_F\equiv \partial\Lned/\partial F$ and
\begin{equation}
 T_{\mu}^{\ \nu}=\frac{1}{4\pi}\left(\Lned_F F_{\mu\lambda}F^{\nu\lambda}
 -\frac{1}{4}\delta_\mu^{\ \nu}\Lned\right).
 \label{eq:stress}
\end{equation}
For a magnetic monopole,
\begin{equation}
 A_\phi=-q\cos\theta,
 \qquad
 F_{\theta\phi}=q\sin\theta,
 \qquad
 F=\frac{2q^2}{r^4} .
 \label{eq:magfield}
\end{equation}
Writing
\begin{equation}
 f(r)=1-\frac{2m(r)}{r},
 \label{eq:massfunctiondef}
\end{equation}
Einstein's equations imply the standard inverse-reconstruction identities \cite{Toshmatov2021}
\begin{equation}
 \Lned(r)=\frac{4m'(r)}{r^2},
 \qquad
 \Lned_F(r)=\frac{r^2\left[2m'(r)-r m''(r)\right]}{2q^2} .
 \label{eq:inverseNED}
\end{equation}
These equations describe the magnetic branch of the Einstein--NED system.
The electric branch may instead be formulated using the dual Hamiltonian
$P$ framework, whereas the magnetic representation directly produces a
single-valued local $\Lned(F)$ with Maxwell asymptotics.
For the extraordinary polarization, the characteristic metric may be
represented as
\begin{equation}
\begin{aligned}
 \dd s_{\rm eff}^2
 &=-\frac{f(r)}{\Lned_F(r)}\dd t^2
 +\frac{\dd r^2}{f(r)\Lned_F(r)}
 +\frac{r^2}{\Phi(r)}\dd\Omega^2,
 \\
 \Phi&=\Lned_F+2F\Lned_{FF}.
\end{aligned}
 \label{eq:effmetric}
\end{equation}
The ordinary characteristic remains on the background null cone.  This
polarization distinction is retained below.

\section{T-duality inspired NED solution}
\label{sec:solution}

The metric \eqref{eq:newmetric} corresponds to the mass function
\begin{equation}
 m(r)=\frac{M r^3}{(r^2+\ell^2)^{3/2}}
      -\frac{q^2 r^3}{2(r^2+\ell^2)^2} .
 \label{eq:massfunction}
\end{equation}
The first term is the mass profile generated by the T-duality smeared Newtonian source, while the second term is the finite NED charge contribution.  The large-radius expansion gives
\begin{equation}
 f(r)=1-\frac{2M}{r}+\frac{q^2}{r^2}
 +\frac{3M\ell^2}{r^3}-\frac{2q^2\ell^2}{r^4}+\mathcal{O}(r^{-5}),
 \label{eq:asymptotic}
\end{equation}
so $M$ is the ADM mass and $q$ is the asymptotic Maxwell charge.  Therefore the solution approaches RN in the infrared, with T-duality corrections suppressed by powers of $\ell/r$ \cite{Nicolini2019}.

Near the origin one obtains
\begin{equation}
 f(r)=1-\left(\frac{2M}{\ell^3}-\frac{q^2}{\ell^4}\right)r^2
 +\mathcal{O}(r^4).
 \label{eq:core}
\end{equation}
The central core is de Sitter for $2M\ell>q^2$, Minkowski for $2M\ell=q^2$, and anti-de Sitter for $2M\ell<q^2$.  The effective core cosmological constant is
\begin{equation}
 \Lambda_{\rm eff}=3\left(\frac{2M}{\ell^3}-\frac{q^2}{\ell^4}\right).
 \label{eq:lambdaeff}
\end{equation}
The condition $2M\ell>q^2$ selects a de Sitter center with positive central energy density. This local condition does not by itself guarantee that the weak energy condition is satisfied at every radius, as shown below.

\subsection{Curvature regularity}

For a line element of the form \eqref{eq:metric}, the Kretschmann scalar is
\begin{equation}
 K=f''(r)^2+4\left(\frac{f'(r)}{r}\right)^2
 +4\left(\frac{1-f(r)}{r^2}\right)^2 .
 \label{eq:kretschmannformula}
\end{equation}
Using Eq.~\eqref{eq:core}, the central values of the curvature invariants are finite:
\begin{align}
 R(0)&=12\left(\frac{2M}{\ell^3}-\frac{q^2}{\ell^4}\right),
 \label{eq:R0}\\
 R_{\mu\nu}R^{\mu\nu}\big|_{r=0}&=36\left(\frac{2M}{\ell^3}-\frac{q^2}{\ell^4}\right)^2,
 \label{eq:Ricci2}\\
 K(0)&=24\left(\frac{2M}{\ell^3}-\frac{q^2}{\ell^4}\right)^2 .
 \label{eq:K0}
\end{align}
Thus the central curvature singularity of RN is removed for every finite nonzero $\ell$.

\subsection{Energy conditions}
\label{subsec:energyconditions}

Using the mass-function definition \eqref{eq:massfunctiondef}, the
effective anisotropic-fluid variables associated with the NED source are
\begin{equation}
 \rho=\frac{m'(r)}{4\pi r^2},
 \qquad
 p_r=-\rho,
 \qquad
 p_t=-\frac{m''(r)}{8\pi r}.
 \label{eq:fluidvariables}
\end{equation}
Hence the weak energy condition requires
\begin{equation}
 \rho\ge0,
 \qquad
 \rho+p_r=0,
 \qquad
 \rho+p_t=\frac{2m'(r)-rm''(r)}{8\pi r^2}\ge0 .
 \label{eq:wecconditions}
\end{equation}
For the present mass function,
\begin{equation}
 m'(r)=
 \frac{r^2\left[
 6M\ell^2\sqrt{r^2+\ell^2}
 -3q^2\ell^2
 +q^2r^2
 \right]}
 {2(r^2+\ell^2)^3},
 \label{eq:mprimewec}
\end{equation}
and
\begin{align}
 2m'(r)-rm''(r)
 &=
 \frac{r^4}{(r^2+\ell^2)^{9/2}}
 \bigg[
 15M\ell^4+15M\ell^2r^2
 \nonumber\\
&
 -10q^2\ell^2\sqrt{r^2+\ell^2}
 +2q^2r^2\sqrt{r^2+\ell^2}
 \bigg].
 \label{eq:wecsecond}
\end{align}
Near the center one finds
\begin{equation}
 \rho(0)=
 \frac{3(2M\ell-q^2)}{8\pi\ell^4}.
 \label{eq:rhozero}
\end{equation}
Thus $\rho(0)\geq0$ is equivalent to
\begin{equation}
 2M\ell\geq q^2.
 \label{eq:centraldensitycondition}
\end{equation}
The global weak energy condition can also be determined analytically.  Set
\begin{equation}
 x=\frac{r}{\ell},
 \qquad
 \mu=\frac{M}{\ell},
 \qquad
 e=\frac{q}{\ell},
 \qquad
 s=\sqrt{1+x^2}\geq1.
 \label{eq:dimensionless_variables}
\end{equation}
The numerator of $m'(r)$ is proportional to
$e^2s^2+6\mu s-4e^2$, which is monotone increasing for $s\geq1$.
It follows that $\rho(r)\geq0$ at every radius if and only if
Eq.~\eqref{eq:centraldensitycondition} holds.  Similarly, apart from an
everywhere nonnegative factor, the numerator of $2m'-rm''$ is
\begin{equation}
 s\left(2e^2s^2+15\mu s-12e^2\right).
 \label{eq:wec_monotone_factor}
\end{equation}
The term in parentheses is also monotone increasing on $s\geq1$, so its
minimum occurs at $s=1$.  Therefore $\rho+p_t\geq0$ globally if and only
if
\begin{equation}
 3\mu\geq2e^2
 \quad\Longleftrightarrow\quad
 3M\ell\geq2q^2.
 \label{eq:globalwec}
\end{equation}
This is stronger than Eq.~\eqref{eq:centraldensitycondition}; hence
Eq.~\eqref{eq:globalwec} is necessary and sufficient for the complete WEC.
A de Sitter center with positive central density alone does not guarantee
global satisfaction of the WEC.

\subsection{NED Lagrangian}

From Eqs.~\eqref{eq:inverseNED} and \eqref{eq:massfunction}, the NED Lagrangian supporting the geometry is
\begin{equation}
\Lned(r)=\frac{2\left[6M\ell^2\sqrt{r^2+\ell^2}
 +q^2(r^2-3\ell^2)\right]}{(r^2+\ell^2)^3}\, .
 \label{eq:Lr}
\end{equation}

More explicitly, the large-radius expansions are \begin{align} \Lned(r) &= \frac{2q^2}{r^4} +\frac{12M\ell^2}{r^5} -\frac{12q^2\ell^2}{r^6} +\mathcal{O}(r^{-7}), \label{eq:Lweakexpanded}\\ \Lned_F(r) &= 1+\frac{15M\ell^2}{2q^2r} -\frac{9\ell^2}{r^2} +\mathcal{O}(r^{-3}). \label{eq:LFweakexpanded} \end{align} The Maxwell limit is therefore recovered continuously. However, the leading correction to $\Lned(F)$ is proportional to $F^{5/4}$, so the reconstructed theory is not analytic in integer powers of $F$ around $F=0$. In particular, higher derivatives such as $\Lned_{FF}$ need not remain finite in the strict vacuum limit, even though $\Lned_F\rightarrow1$.
The derivative with respect to the invariant is
\begin{equation}
\begin{aligned}
 \Lned_F(r)&=\frac{r^6}{2q^2(r^2+\ell^2)^{9/2}}
 \big[15M\ell^4+15M\ell^2 r^2
 \\
&\quad -10q^2\ell^2\sqrt{r^2+\ell^2}
 +2q^2r^2\sqrt{r^2+\ell^2}\big] .
\end{aligned}
 \label{eq:LFr}
\end{equation}
The magnetic invariant is $F=2q^2/r^4$, so Eq.~\eqref{eq:Lr} can be written explicitly as a function of $F$.  It is convenient to define
\begin{equation}
 y=\frac{\ell}{r}=\left(\frac{F}{F_\ell}\right)^{1/4},
 \qquad
 F_\ell=\frac{2q^2}{\ell^4},
 \label{eq:ned_field_variable}
\end{equation}
where $\mu$ and $e$ are defined in
Eq.~\eqref{eq:dimensionless_variables}.
Then
\begin{equation}
\begin{aligned}
 \Lned(F)&=\frac{2}{\ell^2}
 \frac{6\mu y^5\sqrt{1+y^2}+e^2y^4(1-3y^2)}{(1+y^2)^3},
 \\
 y&=\left(\frac{F\ell^4}{2q^2}\right)^{1/4} .
\end{aligned}
 \label{eq:LFexplicit}
\end{equation}

At strong field, $r\to0$ corresponds to $F\to\infty$. The reconstructed magnetic Lagrangian and its first derivative have the limits \begin{equation} \lim_{r\to0}\Lned(r) = \lim_{F\to\infty}\Lned(F) = \frac{12M}{\ell^3}-\frac{6q^2}{\ell^4} = \frac{6(2M\ell-q^2)}{\ell^4}, \label{eq:Lstrongfield} \end{equation} and \begin{equation} \lim_{r\to0}\Lned_F(r) = \lim_{F\to\infty}\Lned_F(F) =0. \label{eq:LFstrongfield} \end{equation} Consequently, the stress tensor remains finite at the center. The tuned case $2M\ell=q^2$ has a locally Minkowski center and satisfies $\lim_{F\to\infty}\Lned(F)=0$. The cases $2M\ell>q^2$ and $2M\ell<q^2$ correspond, respectively, to de Sitter and anti-de Sitter central geometries.

The reconstructed function \eqref{eq:LFexplicit} depends explicitly on the parameters $M$, $q$, and $\ell$. This feature has an important interpretational consequence. In a conventional Einstein--NED model, the electromagnetic Lagrangian is fixed before the field equations are solved, while the ADM mass and magnetic charge arise as integration constants. In the present inverse construction, changing $M$ or $q$ generally changes the reconstructed matter Lagrangian itself. Consequently, the resulting family should be regarded as a parameter-dependent effective NED representation of the prescribed geometry. It does not, without an additional microscopic construction, define a continuous family of states belonging to one universal local electromagnetic theory.

\section{Horizons and extremality}
\label{sec:horizons}

Using the dimensionless variables in
Eq.~\eqref{eq:dimensionless_variables}, the metric function is
\begin{equation}
 f(x)=1-\frac{2\mu x^2}{(1+x^2)^{3/2}}
 +\frac{e^2 x^2}{(1+x^2)^2} .
 \label{eq:fdimless}
\end{equation}
The horizon equation $f(x_+)=0$ gives the mass as a function of the horizon radius,
\begin{equation}
 \mu_+(x_+,e)=\frac{e^2x_+^2+(1+x_+^2)^2}{2x_+^2\sqrt{1+x_+^2}} .
 \label{eq:muofx}
\end{equation}
Extremality is defined by $f(x_e)=f'(x_e)=0$.  Eliminating $\mu$ yields
\begin{equation} x_e^6-e^2x_e^4-3x_e^2-2=0. \label{eq:extpoly} \end{equation}
For $e=0$, Eq.~\eqref{eq:extpoly} gives $x_e=\sqrt{2}$ and
\begin{equation}
 \mu_e(0)=\frac{3\sqrt{3}}{4},
\end{equation}
which is the known neutral T-dual remnant.  For fixed $\ell$, the positive
extremal root increases with $|e|$; the sign of the magnetic charge is
irrelevant to the background geometry.  The causal structure is RN-like:
for $M>M_{\rm ext}$ the spacetime has an outer event horizon and an inner
Cauchy horizon, for $M=M_{\rm ext}$ the two horizons merge, and for
$M<M_{\rm ext}$ the geometry is horizonless but regular.

\section{Thermodynamics}
\label{sec:thermo}

A useful approach to charged-black-hole thermodynamics is to combine the
first law, a non-extremality parameter, and the scaling properties of the
thermodynamic variables. In scale-free Einstein--Maxwell--dilaton-type
theories, the non-extremality parameter obtained from the asymptotic
long-range force can coincide with the horizon quantity $2T_HS$ \cite{Yang:2025rud}.
The present system requires a generalization of this construction,
because the zero-point length $\ell$ is an independent dimensionful
coupling and the reconstructed NED sector depends explicitly on the
black-hole parameters.

\subsection{Horizon mass, temperature, and entropy}

The outer horizon $r_+$ is determined by
\begin{equation}
 f(r_+)=0.
\end{equation}
Solving this equation for the ADM mass gives
\begin{equation}
 M(r_+,q,\ell)
 =
 \frac{
 \left(r_+^2+\ell^2\right)^2+q^2r_+^2
 }{
 2r_+^2\sqrt{r_+^2+\ell^2}
 }.
 \label{eq:thermomass}
\end{equation}
It is convenient to define
\begin{equation}
 {\cal R}_+\equiv\sqrt{r_+^2+\ell^2}.
 \label{eq:Rplus}
\end{equation}
Then
\begin{equation}
 M
 =
 \frac{{\cal R}_+^4+q^2r_+^2}
 {2r_+^2{\cal R}_+}.
\end{equation}

The physical Hawking temperature follows from the surface gravity:
\begin{equation}
 T_H
 =
 \frac{f'(r_+)}{4\pi}.
\end{equation}
Using the horizon relation to eliminate $M$, one obtains
\begin{equation}
 T_H
 =
 \frac{
 r_+^6-q^2r_+^4-3\ell^4r_+^2-2\ell^6
 }{
 4\pi r_+\left(r_+^2+\ell^2\right)^3
 }.
 \label{eq:THphysical}
\end{equation}
In the dimensionless variables of
Eq.~\eqref{eq:dimensionless_variables}, this becomes
\begin{equation}
 T_H
 =
 \frac{
 x_+^6-e^2x_+^4-3x_+^2-2
 }{
 4\pi\ell x_+\left(1+x_+^2\right)^3
 }.
 \label{eq:THdimensionless}
\end{equation}
The extremal condition $T_H=0$ therefore reproduces
Eq.~\eqref{eq:extpoly}.

Since the gravitational Lagrangian is the Einstein--Hilbert Lagrangian
and contains no higher-curvature terms, the Wald entropy is the
Bekenstein--Hawking entropy,
\begin{equation}
 S_{\rm BH}
 =
 \frac{A_+}{4}
 =
 \pi r_+^2.
 \label{eq:Waldareaentropy}
\end{equation}
The entropy is therefore not obtained by naively integrating
$\dd M/T_H$, because varying $M$ changes the effective inverse-reconstructed
NED source.

For $r_+\gg\ell$, the temperature has the expansion
\begin{equation}
 T_H
 =
 \frac{1}{4\pi r_+}
 \left[
 1-\frac{q^2+3\ell^2}{r_+^2}
 +\frac{3q^2\ell^2+3\ell^4}{r_+^4}
 +\mathcal{O}(r_+^{-6})
 \right].
 \label{eq:THlarge}
\end{equation}
The Reissner--Nordstr\"om result is recovered for $\ell\to0$, whereas
the terms proportional to $\ell$ encode the leading zero-point-length
corrections.

\subsection{Asymptotic and horizon non-extremality}

Because the spacetime is asymptotically Reissner--Nordstr\"om, the
leading long-range force between two identical objects is controlled by
\begin{equation}
 q^2-M^2.
\end{equation}
One may therefore define an asymptotic-force parameter
\begin{equation}
 \mu_\infty^2=M^2-q^2.
 \label{eq:asymptoticnonextremality}
\end{equation}
In ordinary Reissner--Nordstr\"om thermodynamics, $\mu_\infty$ vanishes
at extremality and is directly related to the horizon non-extremality.

This identification does not hold for the present regular black hole.
For example, the neutral extremal solution has $q=0$ and
\begin{equation}
 M_{\rm ext}=\frac{3\sqrt{3}}{4}\ell,
\end{equation}
so that $\mu_\infty=M_{\rm ext}\neq0$ even though $T_H=0$. Horizon
extremality is therefore not equivalent to cancellation of the
asymptotic gravitational and electromagnetic forces.

The appropriate horizon non-extremality parameter is instead
\begin{equation}
 {\cal N}_H
 \equiv
 2T_HS_{\rm BH}.
 \label{eq:horizonnonextremalitydef}
\end{equation}
For the present solution,
\begin{equation}
 {\cal N}_H
 =
 \frac{
 r_+\left(
 r_+^6-q^2r_+^4-3\ell^4r_+^2-2\ell^6
 \right)
 }{
 2\left(r_+^2+\ell^2\right)^3
 }.
 \label{eq:horizonnonextremality}
\end{equation}
Equivalently,
\begin{equation}
 \frac{{\cal N}_H}{\ell}
 =
 \frac{
 x_+\left(
 x_+^6-e^2x_+^4-3x_+^2-2
 \right)
 }{
 2\left(1+x_+^2\right)^3
 }.
 \label{eq:horizonnonextremalitydimensionless}
\end{equation}
This quantity vanishes precisely at the degenerate horizon and is
positive on the nonextremal outer-horizon branch.

The inequivalence
\begin{equation}
 \mu_\infty\neq{\cal N}_H
\end{equation}
is a direct consequence of the additional scale $\ell$ and shows why
the scale-free asymptotic--horizon relation cannot be applied unchanged
to this geometry.

\subsection{Homogeneity and extended Smarr relations}

Although the theory contains the length scale $\ell$, the horizon mass
is homogeneous when $\ell$ is allowed to scale together with the other
length variables:
\begin{equation}
 M(\lambda^2S_{\rm BH},\lambda q,\lambda\ell)
 =
 \lambda M(S_{\rm BH},q,\ell).
 \label{eq:masshomogeneity}
\end{equation}
Euler's theorem consequently gives
\begin{equation}
 M
 =
 2S_{\rm BH}
 \left(\frac{\partial M}{\partial S_{\rm BH}}\right)_{q,\ell}
 +
 q\left(\frac{\partial M}{\partial q}\right)_{S_{\rm BH},\ell}
 +
 \ell\left(\frac{\partial M}{\partial\ell}\right)_{S_{\rm BH},q}.
 \label{eq:EulerSmarr}
\end{equation}

Direct differentiation of Eq.~\eqref{eq:thermomass} gives
\begin{equation}
 \left(\frac{\partial M}{\partial q}\right)_{S_{\rm BH},\ell}
 =
 \frac{q}{{\cal R}_+},
 \label{eq:phiqbare}
\end{equation}
and
\begin{equation}
 \left(\frac{\partial M}{\partial\ell}\right)_{S_{\rm BH},q}
 =
 \frac{
 \ell\left(3{\cal R}_+^4-q^2r_+^2\right)
 }{
 2r_+^2{\cal R}_+^3
 }.
 \label{eq:psilbare}
\end{equation}
The entropy derivative of the horizon mass is related to the Hawking
temperature by
\begin{equation}
 T_H
 =
 \Xi_+
 \left(\frac{\partial M}{\partial S_{\rm BH}}\right)_{q,\ell},
 \label{eq:TrelationXi}
\end{equation}
where
\begin{equation}
 \Xi_+
 =
 \frac{r_+^3}{\left(r_+^2+\ell^2\right)^{3/2}}
 =
 \frac{x_+^3}{\left(1+x_+^2\right)^{3/2}}.
 \label{eq:Xiplus}
\end{equation}
Thus the exact differential of the horizon mass can be written as
\begin{equation}
 \dd M
 =
 T_{\rm th}\dd S_{\rm BH}
 +\phi_q\dd q
 +\psi_\ell\dd\ell,
 \label{eq:exactstatedifferential}
\end{equation}
where
\begin{equation}
 T_{\rm th}
 =
 \frac{T_H}{\Xi_+},
 \qquad
 \phi_q
 =
 \frac{q}{{\cal R}_+},
 \qquad
 \psi_\ell
 =
 \frac{
 \ell\left(3{\cal R}_+^4-q^2r_+^2\right)
 }{
 2r_+^2{\cal R}_+^3
 }.
 \label{eq:effectiveconjugates}
\end{equation}
Here $T_{\rm th}$ is the derivative of the horizon mass with respect to
the area entropy. It should not be confused with the physical Hawking
temperature measured at infinity. Their difference reflects the
explicit dependence of the inverse-reconstructed matter sector on the
mass parameter.

Equation \eqref{eq:exactstatedifferential} leads to the homogeneous
state-space Smarr relation
\begin{equation}
 M
 =
 2T_{\rm th}S_{\rm BH}
 +\phi_q q
 +\psi_\ell\ell.
 \label{eq:effectiveSmarr}
\end{equation}

Alternatively, multiplying the horizon variation by $\Xi_+$ gives
\begin{equation}
 \Xi_+\dd M
 =
 T_H\dd S_{\rm BH}
 +\Phi_q\dd q
 +\Pi_\ell\dd\ell,
 \label{eq:modifiedfirstlawfull}
\end{equation}
with
\begin{equation}
 \Phi_q
 =
 \Xi_+\phi_q
 =
 \frac{qr_+^3}
 {\left(r_+^2+\ell^2\right)^2},
 \label{eq:modifiedchargepotential}
\end{equation}
and
\begin{equation}
 \Pi_\ell
 =
 \Xi_+\psi_\ell
 =
 \frac{
 \ell r_+
 \left[
 3\left(r_+^2+\ell^2\right)^2-q^2r_+^2
 \right]
 }{
 2\left(r_+^2+\ell^2\right)^3
 }.
 \label{eq:modifiedellpotential}
\end{equation}
The corresponding physical-temperature Smarr identity is
\begin{equation}
\Xi_+M
 =
 2T_HS_{\rm BH}
 +\Phi_q q
 +\Pi_\ell\ell.
 \label{eq:modifiedSmarr}
\end{equation}
Equation \eqref{eq:modifiedSmarr} is the appropriate extension of the
asymptotic--horizon thermodynamic relation to the present two-scale
inverse-NED geometry. In the Reissner--Nordstr\"om limit
$\ell\to0$, one has $\Xi_+\to1$, $\Pi_\ell\to0$, and
$\Phi_q\to q/r_+$, so the standard RN Smarr relation is recovered.

The factor $\Xi_+$ should not be interpreted as a correction to the
Wald entropy. Instead, it accounts for the fact that variations of
$M$ change the effective matter Lagrangian supporting the geometry.

\subsection{Heat capacity and local thermodynamic stability}

At fixed $(q,\ell)$, define
\begin{equation}
 {\cal P}_6(r_+)
 =
 r_+^6-q^2r_+^4-3\ell^4r_+^2-2\ell^6.
 \label{eq:P6physical}
\end{equation}
The positive-temperature outer-horizon branch satisfies
\begin{equation}
 {\cal P}_6(r_+)>0.
\end{equation}
Differentiating the Hawking temperature gives
\begin{equation}
 \left(\frac{\partial T_H}{\partial r_+}\right)_{q,\ell}
 =
 \frac{{\cal D}_8(r_+)}
 {4\pi r_+^2\left(r_+^2+\ell^2\right)^4},
 \label{eq:dTdr}
\end{equation}
where
\begin{equation}
\begin{aligned}
 {\cal D}_8(r_+)
 ={}&
 2\ell^8+11\ell^6r_+^2
 +15\ell^4r_+^4
 -3\ell^2q^2r_+^4
 \\
 &+
 5\ell^2r_+^6
 +3q^2r_+^6-r_+^8.
 \label{eq:D8physical}
\end{aligned}
\end{equation}

The heat capacity associated with the physical Hawking temperature and
the Wald entropy is
\begin{equation}
 C_{q,\ell}
 =
 T_H
 \left(\frac{\partial S_{\rm BH}}{\partial T_H}\right)_{q,\ell}.
 \label{eq:heatcapacitydefinition}
\end{equation}
Its exact expression is
\begin{equation}
 C_{q,\ell}
 =
 \frac{
 2\pi r_+^2\left(r_+^2+\ell^2\right)
 {\cal P}_6(r_+)
 }{
 {\cal D}_8(r_+)
 }.
 \label{eq:heatcapacityexplicit}
\end{equation}
In dimensionless variables,
\begin{equation}
 C_{q,\ell}
 =
 2\pi\ell^2
 \frac{
 x_+^2\left(1+x_+^2\right)
 \left(x_+^6-e^2x_+^4-3x_+^2-2\right)
 }{
 {\cal D}(x_+,e)
 },
 \label{eq:heatcapacitydimensionless}
\end{equation}
where
\begin{equation}
\begin{aligned}
 {\cal D}(x,e)
 ={}&
 2+11x^2
 +(15-3e^2)x^4
 \\
 &+
 (5+3e^2)x^6-x^8.
 \label{eq:Ddimensionless}
\end{aligned}
\end{equation}

The extremal configuration satisfies
\begin{equation}
 C_{q,\ell}=0.
\end{equation}
On the positive-temperature branch, the sign of the heat capacity is
determined by ${\cal D}(x_+,e)$. Hence
\begin{equation}
 {\cal D}(x_+,e)>0
\end{equation}
corresponds to local canonical stability, while
\begin{equation}
 {\cal D}(x_+,e)<0
\end{equation}
corresponds to an unstable branch. The equation
\begin{equation}
 {\cal D}(x_+,e)=0
 \label{eq:Daviescondition}
\end{equation}
determines the stationary points of the Hawking temperature and the
divergences of the heat capacity. These points are Davies-type
thermodynamic transition points. Since the matter theory is
parameter-dependent, they should be interpreted as transitions in the
effective horizon thermodynamics rather than phase transitions of one
fixed microscopic NED theory.

The derivative entering the specific-heat bounds is
\begin{equation}
 U
 \equiv
 \left(
 \frac{\partial T_H}{\partial S_{\rm BH}}
 \right)_{q,\ell}
 =
 \frac{
 {\cal D}_8(r_+)
 }{
 8\pi^2r_+^3
 \left(r_+^2+\ell^2\right)^4
 }.
 \label{eq:Uthermo}
\end{equation}

\subsection{Thermodynamic inequalities}

The Penrose-type entropy quantity is
\begin{equation}
 Y
 \equiv
 M-\frac{1}{2}\sqrt{\frac{S_{\rm BH}}{\pi}}
 =
 M-\frac{r_+}{2}.
 \label{eq:Ydefinition}
\end{equation}
For the present solution,
\begin{equation}
 Y
 =
 \frac{
 {\cal R}_+\left({\cal R}_+^3-r_+^3\right)
 +q^2r_+^2
 }{
 2r_+^2{\cal R}_+
 }.
 \label{eq:Yexplicit}
\end{equation}
Since ${\cal R}_+\geq r_+$, it follows immediately that
\begin{equation}
 Y\geq0.
 \label{eq:Ybound}
\end{equation}
Equality occurs only in the simultaneous Schwarzschild limit
$q=\ell=0$.

A stronger quantity proposed for scale-free static black holes is
\begin{equation}
 Z
 \equiv
 M+2T_HS_{\rm BH}
 -\sqrt{\frac{S_{\rm BH}}{\pi}}
 =
 M+{\cal N}_H-r_+.
 \label{eq:Zdefinition}
\end{equation}
For the present solution, its large-horizon expansion is
\begin{equation}
 Z
 =
 -\frac{3\ell^2}{4r_+}
 +
 \frac{
 27\ell^4+20\ell^2q^2
 }{
 16r_+^3
 }
 +\mathcal{O}(r_+^{-5}).
 \label{eq:Zasymptotic}
\end{equation}
Thus $Z$ is negative for sufficiently large $r_+$ whenever
$\ell\neq0$. The bound $Z\geq0$ is therefore not universal in the
presence of the independent zero-point length. This violation is
consistent with the fact that the derivation of the scale-free bound
does not include an additional dimensionful coupling such as $\ell$.
The specific-heat inequalities
\begin{equation}
 -\frac{1}{2}\frac{T_H}{S_{\rm BH}}
 \leq
 U
 \leq
 \frac{1}{4\sqrt{\pi}}S_{\rm BH}^{-3/2}
 \label{eq:specificheatbounds}
\end{equation}
can nevertheless be checked analytically. The difference from the lower
bound is
\begin{equation}
\begin{aligned}
 U+\frac{T_H}{2S_{\rm BH}}
 =
 \frac{
 3\ell^2\left(r_+^2+\ell^2\right)^2
 +q^2r_+^2\left(r_+^2-2\ell^2\right)
 }{
 4\pi^2r_+
 \left(r_+^2+\ell^2\right)^4
 }.
 \label{eq:lowerbounddifference}
\end{aligned}
\end{equation}
The difference from the upper bound is
\begin{equation}
\begin{aligned}
 \frac{1}{4\sqrt{\pi}}S_{\rm BH}^{-3/2}-U
 =
 \frac{
 3\left(r_+^2-\ell^2\right)
 \left[
 \left(r_+^2+\ell^2\right)^2-q^2r_+^2
 \right]
 }{
 8\pi^2r_+
 \left(r_+^2+\ell^2\right)^4
 }.
 \label{eq:upperbounddifference}
\end{aligned}
\end{equation}
The outer black-hole branch satisfies
\begin{equation}
 r_+\geq r_e\geq\sqrt{2}\,\ell,
\end{equation}
and
\begin{equation}
 \left(r_+^2+\ell^2\right)^2-q^2r_+^2>0.
\end{equation}
Consequently, both right-hand sides of
Eqs.~\eqref{eq:lowerbounddifference} and
\eqref{eq:upperbounddifference} are nonnegative, and the specific-heat
bounds are satisfied throughout the outer-horizon branch.

The thermodynamic analysis therefore exhibits a mixed pattern. The
Penrose entropy inequality and the specific-heat bounds remain valid,
whereas the stronger scale-free quantity $Z$ is generally violated by
the zero-point-length corrections. This provides a quantitative
thermodynamic signature of the additional T-duality-inspired scale.

\section{Physical parameter domain and optical admissibility}
\label{sec:physical_parameter_map}

Regularity of the background geometry does not, in general, guarantee that
the characteristic geometry of nonlinear electrodynamics (NED) remains
nondegenerate outside the event horizon.  For the present solution, however,
the horizon, energy-condition, and optical requirements can be organized
almost entirely analytically.  We use the variables in
Eq.~\eqref{eq:dimensionless_variables} and the metric function
\eqref{eq:fdimless}.
The background depends on the magnetic charge only through $e^2$; hence it
is sufficient to display the half-plane $e\geq0$.

\subsection{Horizons and the WEC boundary}

The horizon mass \eqref{eq:muofx} diverges for both
$x_h\rightarrow0^+$ and $x_h\rightarrow\infty$.  With $z=x_h^2$, the
sign of its derivative is controlled by
\begin{equation}
 g(z;e)=z^3-e^2z^2-3z-2.
 \label{eq:map_extremal_polynomial}
\end{equation}
Descartes' rule of signs shows that $g$ has exactly one positive root.
Consequently, the horizon mass possesses a unique global minimum, and the
extremal radius $z_e=x_e^2$ is the positive root of
Eq.~\eqref{eq:extpoly}.
An especially useful parametric representation of the extremality curve is
\begin{align}
 e_{\rm ext}^2(z)
 &=\frac{(z-2)(z+1)^2}{z^2},
 \nonumber\\
 \mu_{\rm ext}(z)
 &=\frac{(z-1)(z+1)^{3/2}}{z^2},
 \qquad z\geq2.
 \label{eq:map_extremal_parametric}
\end{align}
Thus $z_e=2$ in the neutral limit and $z_e>2$ for $e>0$.  The horizon
classification stated in Sec.~\ref{sec:horizons} is therefore global.
When horizons exist, $x_+$ denotes the largest positive root of $f=0$.

The global WEC boundary follows from Eq.~\eqref{eq:globalwec}; accordingly,
\begin{equation}
 {\cal D}_{\rm BH+WEC}
 =\left\{(e,\mu):e\geq0,\quad
 \mu\geq\max\!\left[\mu_{\rm ext}(e),\frac{2e^2}{3}\right]
 \right\}.
 \label{eq:map_bh_wec_domain}
\end{equation}
The two lower boundaries intersect at
\begin{align}
 z_*&=4.023454646\ldots,
 &e_*&=1.776029245\ldots,
 \nonumber\\
 \mu_*&=2.102853252\ldots,
 \label{eq:map_crossover_values}
\end{align}
where $z_*>2$ is the physical root of
\begin{equation}
 4z_*^3-21z_*^2+18z_*+7=0.
 \label{eq:map_crossover_cubic}
\end{equation}
Hence extremality supplies the active lower boundary for $0\leq e\leq e_*$,
whereas the WEC supplies it for $e\geq e_*$.

\begin{figure}[t]
 \centering
 \includegraphics[width=\columnwidth]{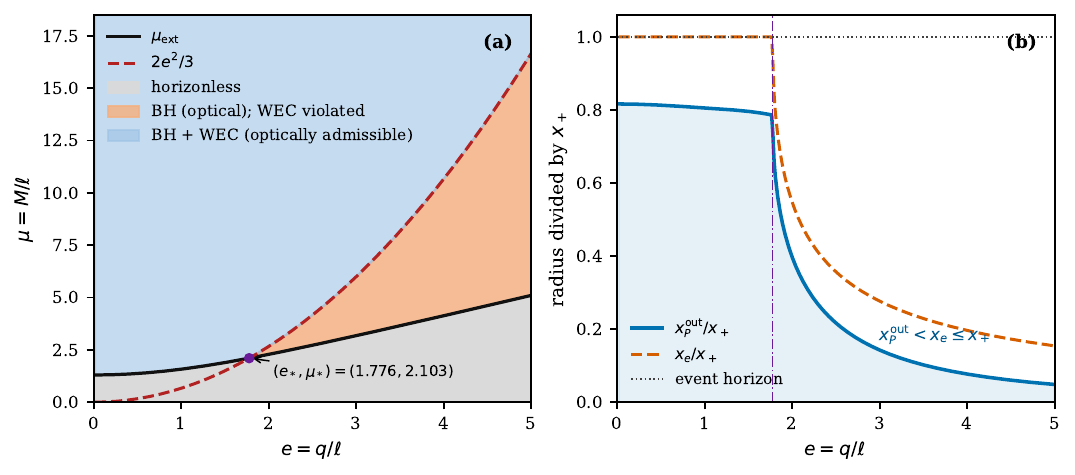}
 \caption{Analytic physical-parameter map.  Panel (a) shows the extremality
 curve, the global-WEC boundary, and their intersection.  Every charged black
 hole above the extremality curve is optically admissible; the blue region is
 the subset that also satisfies the global WEC.  Panel (b) evaluates the
 outermost zero $x_P^{\rm out}$ of $P=\Phi$ along the lower boundary of the
 blue region.  The strict inequality $x_P^{\rm out}<x_e\leq x_+$ confirms
 that the optical degeneracy is hidden behind the event horizon.}
 \label{fig:td_ned_physical_domain}
\end{figure}

\subsection{Analytic optical-admissibility theorem}

For $e\neq0$, define the two characteristic functions
\begin{equation}
 H(x)\equiv{\cal L}_F(x),
 \qquad
 P(x)\equiv\Phi(x)={\cal L}_F(x)+2F{\cal L}_{FF}(x).
 \label{eq:map_optical_functions}
\end{equation}
Since $F\propto x^{-4}$, they obey the exact identity
\begin{equation}
 P(x)=H(x)-\frac{x}{2}\frac{\dd H}{\dd x}.
 \label{eq:map_P_derivative_identity}
\end{equation}
Direct substitution gives
\begin{equation}
 \begin{split}
 H(x)=\frac{x^6}{2e^2(1+x^2)^{9/2}}
 \Big[&15\mu(1+x^2)
 \\
 &+2e^2(x^2-5)\sqrt{1+x^2}\Big],
 \end{split}
 \label{eq:map_H_function}
\end{equation}
and
\begin{equation}
 \begin{split}
 P(x)={}&
 \frac{15\mu x^6(3x^4-x^2-4)}
 {4e^2(1+x^2)^{11/2}}
 \\
 &+\frac{x^6(x^4-13x^2+10)}{(1+x^2)^5}.
 \end{split}
 \label{eq:map_P_function}
\end{equation}
The corresponding representative of the effective characteristic metric is
\begin{equation}
 \begin{split}
 \frac{\dd s_{\rm opt}^2}{\ell^2}=-\frac{f}{H}\dd\tau^2
+\frac{\dd x^2}{fH}+\frac{x^2}{P}\dd\Omega^2,
 \\
 \tau&=\frac{t}{\ell}.
 \end{split}
 \label{eq:map_effective_metric}
\end{equation}
Its determinant is proportional to $-1/(H^2P^2)$.  Because $H,P\to1$ at
infinity, remaining on the nondegenerate branch continuously connected to
Maxwell theory is equivalent to requiring $H>0$ and $P>0$ throughout the
domain of outer communication.

These inequalities can be established analytically.  Let
$s=\sqrt{1+x^2}$.  Equations~\eqref{eq:map_H_function} and
\eqref{eq:map_P_function} become
\begin{equation}
 \begin{aligned}
 H(x)&=\frac{x^6}{2e^2s^8}Q_H(s),\\
 Q_H(s)&=2e^2s^2+15\mu s-12e^2.
 \end{aligned}
 \label{eq:map_QH}
\end{equation}
and
\begin{equation}
 \begin{aligned}
 P(x)&=\frac{x^6}{4e^2s^{10}}Q_P(s),\\
 Q_P(s)&=4e^2(s^4-15s^2+24)
 +15\mu s(3s^2-7).
 \end{aligned}
 \label{eq:map_QP}
\end{equation}
At extremality, Eq.~\eqref{eq:map_extremal_parametric} yields
\begin{equation}
 Q_H(s_e)
 =\frac{(z_e+1)^2}{z_e^2}
 \left(2z_e^2+z_e+5\right)>0.
 \label{eq:map_QH_extremal}
\end{equation}
Moreover, $Q_H'(s)=4e^2s+15\mu>0$, and increasing $\mu$ only increases
$Q_H$.  It follows that
\begin{equation}
 H(x)>0,
 \qquad x\geq x_e(e),
 \qquad \mu\geq\mu_{\rm ext}(e).
 \label{eq:map_H_exterior_theorem}
\end{equation}

For the second characteristic function, its value at the extremal horizon can
be written as
\begin{equation}
 P(x_e)=
 \frac{z_e^3\left(4z_e^3-15z_e^2+39z_e-20\right)}
 {4(z_e-2)(z_e+1)^5}>0,
 \qquad e>0.
 \label{eq:map_P_extremal}
\end{equation}
Indeed, the cubic in parentheses equals $30$ at $z_e=2$ and has the
strictly positive derivative $12z_e^2-30z_e+39$.  In addition,
\begin{equation}
 Q_P'(s_e)
 =\frac{s_e^3}{z_e^2}
 \left(16z_e^3+15z_e^2-33z_e+178\right)>0,
 \label{eq:map_QPprime_extremal}
\end{equation}
while
\begin{equation}
 Q_P''(s)=48e^2s^2+270\mu s-120e^2>0,
 \qquad s\geq s_e\geq\sqrt{3}.
 \label{eq:map_QPsecond_exterior}
\end{equation}
Thus $Q_P$ is positive and increasing for $x\geq x_e$ on the extremal
solution.  For $\mu>\mu_{\rm ext}$ this conclusion is strengthened by
\begin{equation}
 \frac{\partial Q_P}{\partial\mu}=15s(3s^2-7)>0,
 \qquad s\geq s_e.
\end{equation}
Consequently,
\begin{equation}H(x)>0,\qquad P(x)>0,
 \qquad x\geq x_e(e),\quad
 \mu\geq\mu_{\rm ext}(e).
 \label{eq:map_optical_positivity_theorem}
\end{equation}
Since $x_+\geq x_e$, every charged black hole in this family is optically
admissible in the exterior.  In particular,
\begin{equation}
 {\cal D}_{\rm opt}^{(\rm BH)}
 ={\cal D}_{\rm BH}\cap\{e>0\},
 \label{eq:map_optical_domain}
\end{equation}
and optical positivity is not an additional numerical cut on the black-hole
parameter space.

For completeness, on the WEC branch $Q_H(1)=5(3\mu-2e^2)\geq0$, so $H$
has no positive zero.  Moreover,
\begin{equation}
 Q_P(1)=20(2e^2-3\mu)\leq0,
 \qquad
 Q_P''(s)\geq108e^2
 \quad(s\geq1).
 \label{eq:map_QP_wec_monotonicity}
\end{equation}
Because $Q_P'$ is strictly increasing and $Q_P\to+\infty$, $P$ has exactly
one noncentral positive zero on this branch.  The near-center expansions are
\begin{align}
 H(x)&=\frac{5(3\mu-2e^2)}{2e^2}x^6
 +\frac{21(4e^2-5\mu)}{4e^2}x^8
 +{\cal O}(x^{10}),
 \label{eq:map_H_center}\\
 P(x)&=-\frac{5(3\mu-2e^2)}{e^2}x^6
 -\frac{63(4e^2-5\mu)}{4e^2}x^8
 +{\cal O}(x^{10}).
 \label{eq:map_P_center}
\end{align}
At the saturated WEC boundary, $3\mu=2e^2$, these reduce to
\begin{equation}
 H(x)=\frac{7}{2}x^8+{\cal O}(x^{10}),
 \qquad
 P(x)=-\frac{21}{2}x^8+{\cal O}(x^{10}).
 \label{eq:map_center_saturated_wec}
\end{equation}
Thus $P$ is negative sufficiently near the regular center, including at WEC
saturation.  Equation~\eqref{eq:map_optical_positivity_theorem} proves that
its noncentral zero satisfies
\begin{equation}
 x_P^{\rm out}<x_e\leq x_+.
 \label{eq:map_hidden_optical_zero}
\end{equation}
The apparent curves obtained from $f=H=0$ or $f=P=0$ therefore describe
intersections with an inner horizon; they never form boundaries of the
exterior optical domain.

\subsection{Global light-ring selection and shadow radius}

In the exterior, define
\begin{equation}
 {\cal B}(x;\mu,e)
 =\frac{x^2H(x;\mu,e)}{f(x;\mu,e)P(x;\mu,e)}.
 \label{eq:map_impact_function}
\end{equation}
This is the squared impact-parameter function.  Optical positivity implies
${\cal B}>0$, and
\begin{equation}
 \lim_{x\to x_+^+}{\cal B}(x)=+\infty,
 \qquad
 {\cal B}(x)\sim x^2\quad(x\to\infty).
 \label{eq:map_B_endpoint_limits}
\end{equation}
Hence ${\cal B}$ necessarily possesses at least one finite global minimum in
the exterior.  The shadow-generating critical impact parameter must be defined
by this global minimum,
\begin{equation}
 \left(\frac{R_{\rm sh}^{\rm NED}}{\ell}\right)^2
 =\min_{x>x_+}{\cal B}(x;\mu,e)
 ={\cal B}(x_{\rm ph}^{\rm NED};\mu,e).
 \label{eq:map_ned_shadow}
\end{equation}
For a unique nondegenerate minimum,
\begin{equation}
 {\cal B}'(x_{\rm ph}^{\rm NED})=0,
 \qquad
 {\cal B}''(x_{\rm ph}^{\rm NED})>0,
 \label{eq:map_ned_light_ring}
\end{equation}
or, equivalently, ${\cal U}_{\rm NED}=1/{\cal B}$ has a strict local maximum.
This formulation is more precise than selecting an unspecified stationary
point: if several extrema occur, the capture separatrix is determined by the
global minimum of ${\cal B}$.  Marginal configurations, if present, satisfy
${\cal B}'={\cal B}''=0$ and should be reported separately rather than
discarded by the optical-admissibility test.

For comparison, the background-geometric critical radius is determined from
\begin{equation}
 \left(\frac{R_{\rm sh}^{\rm geo}}{\ell}\right)^2
 =\min_{x>x_+}\frac{x^2}{f(x)},
 \qquad
 2f(x_{\rm ph}^{\rm geo})
 -x_{\rm ph}^{\rm geo}f'(x_{\rm ph}^{\rm geo})=0.
 \label{eq:map_geometric_shadow}
\end{equation}
The relative optical deformation is then
\begin{equation}
 \Delta_{\rm opt}(\mu,e)
 =\frac{R_{\rm sh}^{\rm NED}-R_{\rm sh}^{\rm geo}}
 {R_{\rm sh}^{\rm geo}}.
 \label{eq:map_relative_shadow_deformation}
\end{equation}
No fixed sign should be assigned to $\Delta_{\rm opt}$ a priori.

Combining the analytic results, the charged black-hole parameter domain that
also satisfies the global WEC is simply
\begin{equation}
 {\cal D}_{\rm phys}
 =\left\{(e,\mu):e>0,\quad
 \mu\geq\max\!\left[\mu_{\rm ext}(e),\frac{2e^2}{3}\right]
 \right\},
 \label{eq:map_complete_physical_domain}
\end{equation}
with a possible marginal-light-ring subset identified separately through
${\cal B}'={\cal B}''=0$.  The neutral axis $e=0$ must still be treated at the
metric level because the geometry has a smooth neutral limit, whereas the
inverse magnetic-NED representation and the functions $H$ and $P$ contain
explicit inverse powers of $e^2$.  Numerically, one therefore needs only to
solve for $x_+$ and minimize ${\cal B}$ globally; a two-dimensional exterior
zero scan of $H$ and $P$ is not required.

\section{Weak-field classical tests}
\label{sec:classical_tests}

The infrared expansion in Eq.~\eqref{eq:asymptotic} permits a direct
comparison with the classical tests of a static gravitational field.  This
comparison requires some care.  The parameter $\ell$ is a length and enters
the metric only through even powers; it is not a dimensionless deformation
of the radial metric coefficient.  Moreover, massive bodies follow timelike
geodesics of the background metric, whereas electromagnetic rays may follow
either background null geodesics (minimally coupled test radiation) or the
extraordinary characteristic cone of the reconstructed NED.  We keep these
two photon sectors distinct throughout.

\subsection{Geodesic hierarchy}
\label{subsec:ct_geodesics}

In the equatorial plane, the conserved specific energy and angular momentum
of a neutral probe are
\begin{equation}
 E=f(r)\dot t,
 \qquad
 {\cal J}=r^2\dot\phi .
 \label{eq:ct_constants}
\end{equation}
The radial first integral is
\begin{equation}
 \dot r^{\,2}
 =E^2-f(r)\left(\kappa+\frac{{\cal J}^2}{r^2}\right),
 \qquad
 \kappa=\begin{cases}1,&\text{timelike},\\0,&\text{null}.
 \end{cases}
 \label{eq:ct_radial_first_integral}
\end{equation}
For timelike motion, setting $u=1/r$ and expanding the exact metric through
first order in $\ell^2$ gives
\begin{align}
 u''+u={}&\frac{M}{{\cal J}^2}
 -\frac{q^2}{{\cal J}^2}u
 \nonumber\\
 &+\left(3M-\frac{9M\ell^2}{2{\cal J}^2}\right)u^2
 -2q^2u^3
 \nonumber\\
 &+\frac{4q^2\ell^2}{{\cal J}^2}u^3
 -\frac{15}{2}M\ell^2u^4
 \nonumber\\
 &+6q^2\ell^2u^5
 +{\cal O}(\ell^4,u^6).
 \label{eq:ct_binet_timelike}
\end{align}
The corresponding background-null equation is
\begin{equation}
 u''+u
 =3Mu^2-2q^2u^3-\frac{15}{2}M\ell^2u^4
 +6q^2\ell^2u^5+{\cal O}(\ell^4,u^6).
 \label{eq:ct_binet_null}
\end{equation}
Equations~\eqref{eq:ct_binet_timelike} and
\eqref{eq:ct_binet_null} display the relevant hierarchy: charge modifies
the Reissner--Nordstr\"om (RN) terms, while the first purely zero-point-length
contribution is proportional to $M\ell^2$.

\subsection{Periapsis advance}
\label{subsec:ct_periapsis}

Let
\begin{equation}
 p=a(1-e_{\rm orb}^2)=\frac{{\cal J}^2}{M},
 \qquad
 u_0=\frac{1}{p}\left(1+e_{\rm orb}\cos\phi\right),
 \label{eq:ct_kepler_orbit}
\end{equation}
where $e_{\rm orb}$ denotes orbital eccentricity and must not be confused
with the charge ratio $e=q/\ell$ used elsewhere.  Extracting the resonant
$\cos\phi$ terms in Eq.~\eqref{eq:ct_binet_timelike} yields the advance per
radial period
\begin{equation}
 \Delta\varpi
 =\frac{6\pi M}{p}
 -\frac{\pi q^2}{Mp}
 -\frac{9\pi\ell^2}{p^2}
 +{\cal R}_{\varpi}
 \label{eq:ct_periapsis_weak}
\end{equation}
with
\begin{equation}
 {\cal R}_{\varpi}
 ={\cal O}\!\left(
 \frac{M^2}{p^2},\frac{q^2}{p^2},
 \frac{M\ell^2}{p^3},
 \frac{q^2\ell^2}{Mp^3},\frac{\ell^4}{p^4}
 \right).
 \label{eq:ct_periapsis_remainder}
\end{equation}
Both displayed corrections reduce the Schwarzschild prograde advance.  The
explicit $q^2\ell^2/r^4$ term generates the direct first-iteration
contribution
\begin{equation}
 \left.\Delta\varpi\right|_{q^2\ell^2,\,\mathrm{direct}}
 =\frac{12\pi q^2\ell^2}{Mp^3}
 \left(1+\frac{e_{\rm orb}^2}{4}\right),
 \label{eq:ct_periapsis_cross_direct}
\end{equation}
but Eq.~\eqref{eq:ct_periapsis_cross_direct} is not by itself a complete
higher-post-Newtonian result; iterations of lower-order terms contribute at
the same order.

For numerical validation we use the exact metric rather than its asymptotic
truncation.  For a nearly circular orbit,
\begin{align}
 \Omega_\phi^2&=\frac{f'}{2r},
 \nonumber\\
 \Omega_r^2&=\frac{1}{2}ff''-(f')^2+\frac{3ff'}{2r},
 \nonumber\\
 \Delta\varpi_{\rm circ}
 &=2\pi\left(\frac{\Omega_\phi}{\Omega_r}-1\right).
 \label{eq:ct_epicyclic_exact}
\end{align}
For an eccentric orbit with turning radii $r_p$ and $r_a$, the exact
quadrature follows from
\begin{align}
 {\cal J}^2&=
 \frac{f(r_a)-f(r_p)}{f(r_p)/r_p^2-f(r_a)/r_a^2},
 \qquad
 E^2=f(r_p)\left(1+\frac{{\cal J}^2}{r_p^2}\right),
 \label{eq:ct_turning_constants}\\
 \Delta\varpi&=
 2\int_{r_p}^{r_a}
 \frac{{\cal J}\,\dd r}
 {r^2\sqrt{E^2-f(r)(1+{\cal J}^2/r^2)}}-2\pi.
 \label{eq:ct_periapsis_exact}
\end{align}
Panel (a) of Fig.~\ref{fig:td_ned_classical_tests} shows that
Eq.~\eqref{eq:ct_periapsis_weak} approaches the exact epicyclic result in
the expected large-radius regime.

\subsection{Light bending: background and NED characteristics}
\label{subsec:ct_bending}

For background-null rays, $b={\cal J}/E$ and the invariant asymptotic
deflection is
\begin{equation}
 \alpha_{\rm geo}
 =2\int_{r_0}^{\infty}
 \frac{b\,\dd r}{r^2\sqrt{1-b^2f(r)/r^2}}-\pi,
 \qquad
 b^2=\frac{r_0^2}{f(r_0)}.
 \label{eq:ct_bending_exact_geo}
\end{equation}
Expansion at fixed asymptotic impact parameter gives the complete result
through fourth post-Minkowskian order,
\begin{align}
 \alpha_{\rm geo}={}&\frac{4M}{b}
 +\frac{\pi(15M^2-3q^2)}{4b^2}
 \nonumber\\
 &+\frac{1}{b^3}\left(
 \frac{128}{3}M^3-16Mq^2-8M\ell^2\right)
 \nonumber\\
 &+\frac{\pi}{64b^4}\left(
 3465M^4-1890M^2q^2+105q^4\right)
 \nonumber\\
 &+\frac{\pi}{64b^4}\left(
 -1260M^2\ell^2+120q^2\ell^2\right)
 +{\cal O}(b^{-5},\ell^4).
 \label{eq:ct_bending_weak}
\end{align}
The leading zero-point-length response is therefore
\begin{equation}
 \delta\alpha_{\ell}
 =-\frac{8M\ell^2}{b^3},
 \label{eq:ct_bending_ell}
\end{equation}
whereas the direct linear response to the explicit
$-2q^2\ell^2/r^4$ coefficient is
$15\pi q^2\ell^2/(8b^4)$.  The latter is only one component of the full
$b^{-4}$ coefficient displayed in Eq.~\eqref{eq:ct_bending_weak}.

\begin{figure*}[t]
 \centering
 \includegraphics[width=0.98\textwidth]{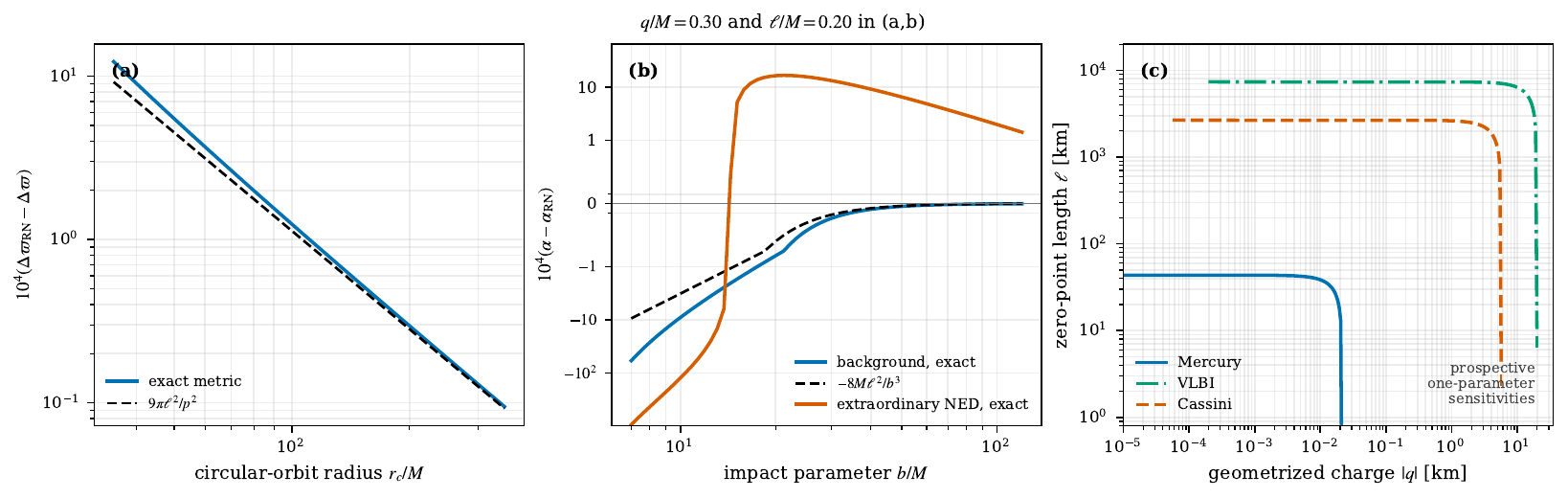}
 \caption{Weak-field signatures of the T-duality-inspired geometry.  (a)
 The exact near-circular zero-point-length residual relative to RN and its
 leading $-9\pi\ell^2/r_c^2$ approximation.  (b) Background-geometric and
 extraordinary-NED deflection residuals relative to RN; the dashed curve is
 the leading geometric term.  The signed vertical scale is symmetric
 logarithmic.  (c) Conditional one-parameter Solar-System sensitivity curves.
 The region below and to the left of a curve is allowed by the corresponding
 proxy.  Panels (a) and (b) use $q/M=0.3$ and $\ell/M=0.2$; all numerical
 trajectories use the exact metric and, for the extraordinary branch, the
 exact functions $\Lned_F$ and $\Phi$.}
 \label{fig:td_ned_classical_tests}
\end{figure*}

The extraordinary NED ray requires a different characteristic cone.  It is
convenient to introduce
\begin{equation}
 \chi(r)=\frac{\Phi(r)}{\Lned_F(r)}.
 \label{eq:ct_chi_definition}
\end{equation}
Then both photon sectors are covered by
\begin{align}
 b^2&=\frac{r_0^2}{f(r_0)\chi(r_0)},
 \label{eq:ct_optical_turning}\\
 \alpha_\chi&=
 2\int_{r_0}^{\infty}
 \frac{b\chi(r)\,\dd r}
 {r^2\sqrt{1-b^2f(r)\chi(r)/r^2}}-\pi,
 \label{eq:ct_optical_bending}
\end{align}
where $\chi=1$ gives Eq.~\eqref{eq:ct_bending_exact_geo}, while
$\chi=\Phi/\Lned_F$ gives the extraordinary branch.  From
Eq.~\eqref{eq:LFweakexpanded},
\begin{equation}
 \chi(r)
 =1+\frac{15M\ell^2}{4q^2r}
 -\frac{9\ell^2}{r^2}
 +{\cal O}\!\left(r^{-3},\ell^4\right).
 \label{eq:ct_chi_asymptotic}
\end{equation}
Although the $1/r$ terms in the individual optical functions are nonzero,
their contribution cancels from the leading $4M/b$ bending coefficient;
the extraordinary and geometric branches first separate at order $b^{-2}$.
Equation~\eqref{eq:ct_chi_asymptotic} is nonuniform as $q\to0$, so the
extraordinary neutral limit cannot be inferred from it.  In addition,
Eqs.~\eqref{eq:ct_optical_turning} and \eqref{eq:ct_optical_bending} are used
only where $\Lned_F>0$ and $\Phi>0$ outside the horizon.  Panel (b) of
Fig.~\ref{fig:td_ned_classical_tests} illustrates the distinction for the
optically admissible point $\ell/M=0.2$ and $q/M=0.3$.

\subsection{Time delay and gravitational redshift}
\label{subsec:ct_delay_redshift}

For a background-null signal with a zeroth-order impact parameter $b$, let
\begin{equation}
 z_i=\sqrt{r_i^2-b^2},
 \qquad
 \vartheta_i=\arccos\!\left(\frac{b}{r_i}\right),
 \qquad i\in\{E,R\}.
 \label{eq:ct_delay_geometry}
\end{equation}
Evaluating the first variation of the optical travel time on the unperturbed
straight line gives
\begin{align}
 K_1(r_i)&=2\ln\!\left(\frac{r_i+z_i}{b}\right)-\frac{z_i}{r_i},
 \nonumber\\
 K_2(r_i)&=\frac{3\vartheta_i}{2b}-\frac{z_i}{2r_i^2},
 \nonumber\\
 K_3(r_i)&=\frac{4z_i}{3b^2r_i}-\frac{z_i}{3r_i^3},
 \nonumber\\
 K_4(r_i)&=\frac{5z_i}{8b^2r_i^2}
 +\frac{5\vartheta_i}{8b^3}-\frac{z_i}{4r_i^4}.
 \label{eq:ct_delay_kernels}
\end{align}
The one-way excess coordinate time for endpoints on opposite sides of the
lens is
\begin{align}
 \Delta t_{\rm 1w}
 ={}&\sum_{i=E,R}\left[
 MK_1(r_i)-\frac{q^2}{2}K_2(r_i)\right.
 \nonumber\\
 &\left.\hspace{1.6cm}
 -\frac{3M\ell^2}{2}K_3(r_i)
 +q^2\ell^2K_4(r_i)
 \right].
 \label{eq:ct_delay_finite}
\end{align}
For $r_E,r_R\gg b$ this reduces to
\begin{equation}
 \Delta t_{\rm 1w}
 \simeq
 2M\left[\ln\!\left(\frac{4r_Er_R}{b^2}\right)-1\right]
 -\frac{3\pi q^2}{4b}
 -\frac{4M\ell^2}{b^2}
 +\frac{5\pi q^2\ell^2}{8b^3}.
 \label{eq:ct_delay_asymptotic}
\end{equation}
The round-trip result is twice Eq.~\eqref{eq:ct_delay_asymptotic}.  The
impact-parameter-independent constant in the Schwarzschild term depends on
the coordinate-delay convention and drops out of a Doppler observable.  In
particular, the leading fractional change in the impact-parameter derivative
is
\begin{equation}
 \left|
 \frac{\delta(\partial_b\Delta t)}
 {(\partial_b\Delta t)_{\rm GR}}
 \right|
 \simeq
 \frac{3\pi q^2}{16Mb}+\frac{2\ell^2}{b^2}.
 \label{eq:ct_delay_fractional}
\end{equation}
For extraordinary rays, the exact coordinate travel time is instead
\begin{equation}
 t_\chi
 =\sum_{i=E,R}\int_{r_0}^{r_i}
 \frac{\dd r}
 {f(r)\sqrt{1-b^2f(r)\chi(r)/r^2}},
 \label{eq:ct_optical_delay}
\end{equation}
with the turning point fixed by Eq.~\eqref{eq:ct_optical_turning}.  This
expression, rather than Eq.~\eqref{eq:ct_delay_asymptotic}, must be used if
the radio signal is identified with the extraordinary NED mode.

A complementary frequency observable follows without orbit integration.
For static emitter and observer,
\begin{equation}
 1+z=\sqrt{\frac{f(r_o)}{f(r_e)}}.
 \label{eq:ct_redshift_exact}
\end{equation}
For an observer at infinity and an emitter at $r$,
\begin{align}
 z_\infty={}&\frac{M}{r}
 +\frac{3M^2-q^2}{2r^2}
 +\frac{5M^3-3Mq^2-3M\ell^2}{2r^3}
 \nonumber\\
 &+\frac{35M^4-30M^2q^2+3q^4}{8r^4}
 \nonumber\\
 &+\frac{-9M^2\ell^2+2q^2\ell^2}{2r^4}
 +{\cal O}(r^{-5},\ell^4).
 \label{eq:ct_redshift_weak}
\end{align}

\subsection{PPN order and prospective sensitivities}
\label{subsec:ct_constraints}

The isotropic-radius transformation begins as
\begin{equation}
 r=\rho+M+\frac{M^2-q^2}{4\rho}
 -\frac{M\ell^2}{2\rho^2}+\cdots .
 \label{eq:ct_isotropic_radius}
\end{equation}
Consequently, the background metric has
\begin{equation}
 \gamma=1,
 \qquad
 \beta_{\rm eff}=1+\frac{q^2}{2M^2},
 \label{eq:ct_ppn_parameters}
\end{equation}
while $\ell$ first enters $g_{tt}$ at order $\rho^{-3}$.  The symbol
$\beta_{\rm eff}$ is only a one-body parametrization of the magnetic hair;
it is not a universal PPN coupling.  In particular, the Cassini result for
$\gamma$ does not directly bound $\ell$.  It can only be used as a proxy for
an unmodeled, impact-parameter-dependent higher-order signal.

To indicate the scale of the weak-field effects, we apply the same
one-parameter error-budget prescription to three measurements, without
claiming a joint statistical fit.  The MESSENGER analysis gives a total
Mercury perihelion uncertainty of $0.0015''$ per century
\cite{Park:2017zgd}.  Using $415.20$ Mercury orbits per century, this
corresponds to
\begin{equation}
 \frac{q^2}{M_\odot p_{\rm Merc}}
 +\frac{9\ell^2}{p_{\rm Merc}^2}
 \lesssim \frac{\sigma_{\varpi}}{\pi}.
 \label{eq:ct_mercury_sensitivity}
\end{equation}
The geodetic-VLBI result $\sigma_\gamma\simeq4.5\times10^{-4}$
\cite{Shapiro:2004zz} is translated using
$\epsilon_{\rm VLBI}=\sigma_\gamma/2$ and $b=R_\odot$.  For Cassini,
$\sigma_\gamma=2.3\times10^{-5}$, the closest ray passed at
$b\simeq1.6R_\odot$, and the corresponding amplitude proxy is
$\epsilon_{\rm Cas}=\sigma_\gamma/2$ \cite{Bertotti:2003rm}.  The last
two estimates use
\begin{equation}
 \frac{3\pi q^2}{16M_\odot b}
 +\frac{2\ell^2}{b^2}\lesssim\epsilon.
 \label{eq:ct_photon_sensitivity}
\end{equation}

\begin{table}[t]
 \caption{Prospective one-standard-deviation sensitivities obtained by
 varying $q$ and $\ell$ one at a time.  The numbers are conditional exterior-
 metric sensitivities, not fitted bounds on the NED black-hole model.}
 \label{tab:ct_sensitivities}
 \centering
 \small
 \begin{ruledtabular}
 \begin{tabular}{lcc}
 Proxy & $|q|_{\max}$ & $\ell_{\max}$ \\
 \hline
 Mercury periapsis & $2.14\times10^{-2}\ {\rm km}$ & $43.7\ {\rm km}$ \\
 Solar VLBI bending & $19.8\ {\rm km}$ & $7.38\times10^3\ {\rm km}$ \\
 Cassini Doppler & $5.66\ {\rm km}$ & $2.67\times10^3\ {\rm km}$
 \end{tabular}
 \end{ruledtabular}
\end{table}

The periapsis proxy is the most sensitive of the three, but even it permits
$\ell$ many orders of magnitude above a microscopic zero-point length.  More
importantly, applying a black-hole exterior to the Sun assumes a matching
source carrying the same magnetic hair and regularization scale.  No such
solar interior has been constructed here.  The numbers in
Table~\ref{tab:ct_sensitivities} therefore quantify the reach of the
observables, not confidence intervals on the fundamental parameters.  A
defensible Solar-System constraint would require inserting the exact force
law into a modern ephemeris, fitting $q^2$ and $\ell^2$ jointly with the solar
quadrupole and planetary initial conditions, and specifying whether the
tracked photons occupy the background or extraordinary characteristic cone.

All limits provide nontrivial checks: $\ell\to0$ reproduces RN,
$q\to0$ gives the neutral T-duality geometry at the level of the background
metric, and $q=\ell=0$ recovers the Schwarzschild classical tests.  The
extraordinary optical formulas, in contrast, require $q\neq0$ and do not
possess a uniform neutral limit.

\section{NED photon sphere and shadow}
\label{sec:shadow}

For a general static, spherically symmetric optical metric,
\begin{equation}
 \dd s_{\rm opt}^2
 =
 -A(r)\dd t^2+B(r)\dd r^2+C(r)\dd\Omega^2,
 \label{eq:generalopticalmetric}
\end{equation}
the shadow depends only on the temporal and angular functions $A(r)$ and
$C(r)$; the radial function $B(r)$ affects the parametrization of the
trajectory but not the location of the circular light ring or the
critical impact parameter
\cite{Vertogradov:2024dpa,Kobialko:2024zhc,Pantig:2025shadow}.

For the extraordinary NED characteristic \eqref{eq:effmetric},
\begin{equation}
 \begin{aligned}
 A(r)&=\frac{f(r)}{\Lned_F(r)},
 &B(r)&=\frac{1}{f(r)\Lned_F(r)},
 \\
 C(r)&=\frac{r^2}{\Phi(r)}.
 \end{aligned}
 \label{eq:ABCned}
\end{equation}
where $\Phi$ is defined in Eq.~\eqref{eq:map_optical_functions}.  Its
radial derivative representation is the dimensional form of
Eq.~\eqref{eq:map_P_derivative_identity}.

Restricting the motion to the equatorial plane, the conserved energy and
angular momentum are $E=A\dot t$ and $L_z=C\dot\phi$.  With
$b=L_z/E$, the null condition gives
\begin{equation}
 A(r)B(r)\dot{r}^{\,2}
 =
 E^2
 \left[
 1-b^2\frac{A(r)}{C(r)}
 \right].
 \label{eq:NEDradialimpact}
\end{equation}

A circular characteristic therefore satisfies
\begin{equation}
 \left.
 \frac{\dd}{\dd r}
 \left(
 \frac{C(r)}{A(r)}
 \right)
 \right|_{r=r_{\rm ph}^{\rm NED}}
 =0,
 \label{eq:NEDphotonconditiongeneral}
\end{equation}
whose logarithmic form is
\begin{equation}
 \left[
 \frac{2}{r}
 +\frac{\Lned_F'}{\Lned_F}
 -\frac{f'}{f}
 -\frac{\Phi'}{\Phi}
 \right]_{r=r_{\rm ph}^{\rm NED}}
 =0.
 \label{eq:NEDphotonconditionlog}
\end{equation}

For the present solution, $C/A=\ell^2{\cal B}$, with ${\cal B}$ given in
Eq.~\eqref{eq:map_impact_function}.  The positivity theorem
\eqref{eq:map_optical_positivity_theorem} guarantees that this function is
regular throughout the charged black-hole exterior.  As shown in
Sec.~\ref{sec:physical_parameter_map}, the capture separatrix is selected
by the global minimum of ${\cal B}$, not by an arbitrarily chosen or merely
outermost stationary point.  In dimensional variables,
\begin{equation}
 b_{\rm c,NED}^2
 =\min_{r>r_+}
 \frac{r^2\Lned_F(r)}{f(r)\Phi(r)}.
 \label{eq:NEDcriticalimpact}
\end{equation}
The asymptotic shadow radius is $R_{\rm sh}^{\rm NED}=b_{\rm c,NED}$,
equivalently Eq.~\eqref{eq:map_ned_shadow}.

If a finite-distance angle is defined using an orthonormal tetrad of the
chosen optical representative, then
\begin{equation}
 \sin^2\Theta_{\rm sh}
 =
 \frac{A(r_{\rm o})}{C(r_{\rm o})}
 b_{\rm c,NED}^2.
 \label{eq:NEDfiniteanglesimple}
\end{equation}
This finite-distance construction is representative-dependent; a physical
angle measured with the background tetrad additionally requires the
constitutive map between the wave covector and the ray velocity.  The
asymptotic result is unambiguous because $f,\Lned_F,\Phi\to1$.

\subsection{Comparison with the background photon sphere}

Background null geodesics are governed by
Eq.~\eqref{eq:map_geometric_shadow}.  Substitution of
Eq.~\eqref{eq:fdimless} gives
\begin{equation}
\begin{aligned}
 &\sqrt{1+x^2}
 \left[
 (1+x^2)^3+2e^2x^4
 \right]
 \\
 &\qquad
 -3\mu x^4(1+x^2)=0.
 \label{eq:geometricphotonexplicit}
\end{aligned}
\end{equation}
In general, the extraordinary and background radii differ because NED
modifies both the temporal and optical-area functions.  Their difference
therefore isolates a propagation effect that is absent from the background
geometry alone.

\subsection{Reissner--Nordstr\"om limit and leading zero-point-length correction}

As a consistency check, consider the limit $\ell\to0$ at fixed nonzero
$q$. In this limit,
\begin{equation}
 \Lned_F\to1,
 \qquad
 \Phi\to1,
\end{equation}
and the effective metric reduces to the Reissner--Nordstr\"om metric. The
outer RN photon-sphere radius is
\begin{equation}
 r_0
 =
 \frac{3M+\sqrt{9M^2-8q^2}}{2},
 \label{eq:RNphotonradius}
\end{equation}
where the existence of the outer circular photon orbit requires
$9M^2\geq8q^2$. Its shadow radius is
\begin{equation}
 \left(R_{\rm sh}^{\rm RN}\right)^2
 =
 \frac{r_0^2}
 {1-2M/r_0+q^2/r_0^2}
 =
 \frac{2r_0^3}{r_0-M}.
 \label{eq:RNshadowradius}
\end{equation}

For a small but finite zero-point length, with
$\ell^2\ll r_0^2$ and fixed $q\neq0$, the expansions of $f$ and
$\Lned_F$ are given by Eqs.~\eqref{eq:asymptotic} and
\eqref{eq:LFweakexpanded}.  Equation~\eqref{eq:map_P_derivative_identity}
then gives
\begin{equation}
 \Phi(r)
 =
 1+\ell^2
 \left(
 \frac{45M}{4q^2r}-\frac{18}{r^2}
 \right)
 +\mathcal{O}(\ell^4).
 \label{eq:Phismalllshadow}
\end{equation}
The NED photon-sphere radius can be written as
\begin{equation}
 r_{\rm ph}^{\rm NED}
 =
 r_0+\ell^2\Delta r_{\rm ph}
 +\mathcal{O}(\ell^4),
 \label{eq:NEDphotonperturbative}
\end{equation}
where
\begin{equation}
 \Delta r_{\rm ph}
 =
 \frac{
 3\left(
 19M^2-15Mr_0+4r_0^2
 \right)
 }{
 8r_0(r_0-3M)(2r_0-3M)
 }.
 \label{eq:NEDphotonshift}
\end{equation}

A useful simplification follows from the fact that the zeroth-order
impact-parameter function is stationary at $r=r_0$. Consequently, the
first-order correction to the shadow radius does not depend explicitly
on the first-order displacement \eqref{eq:NEDphotonshift}. This is the
massless limit of the general perturbative shadow result: translating
the circular orbit contributes only beyond first order because the
shadow function is already extremized at the background photon sphere.

The resulting NED shadow radius is
\begin{equation}
\begin{aligned}
 \left(R_{\rm sh}^{\rm NED}\right)^2
 &=
 \frac{2r_0^3}{r_0-M}
 \bigg[
 1+\ell^2
 \bigg(
 -\frac{15M}{4q^2r_0}
 +\frac{9}{r_0^2}
 \\
 &\qquad\qquad\qquad
 -\frac{2}{r_0(r_0-M)}
 \bigg)
 +\mathcal{O}(\ell^4)
 \bigg].
 \label{eq:NEDshadowperturbative}
\end{aligned}
\end{equation}
For comparison, the shadow calculated from null geodesics of the
background metric alone is
\begin{equation}
 \left(R_{\rm sh}^{\rm geo}\right)^2
 =
 \frac{2r_0^3}{r_0-M}
 \left[
 1-\frac{2\ell^2}{r_0(r_0-M)}
 +\mathcal{O}(\ell^4)
 \right].
 \label{eq:geometricshadowperturbative}
\end{equation}
The additional NED propagation correction is therefore
\begin{equation}
\begin{aligned}
 \left(R_{\rm sh}^{\rm NED}\right)^2
 -
 \left(R_{\rm sh}^{\rm geo}\right)^2
 &=
 \frac{2r_0^3}{r_0-M}\ell^2
 \left(
 -\frac{15M}{4q^2r_0}
 +\frac{9}{r_0^2}
 \right)
 \\
 &\quad
 +\mathcal{O}(\ell^4).
 \label{eq:NEDminusgeometricshadow}
\end{aligned}
\end{equation}
This expression separates the modification of the background geometry
from the genuinely nonlinear-electrodynamic modification of photon
propagation.

The perturbative expressions
\eqref{eq:NEDphotonshift}--\eqref{eq:NEDminusgeometricshadow} are valid
for fixed nonzero $q$ and cannot be continued uniformly to $q=0$. This
is consistent with the fact that the inverse magnetic NED
representation itself is defined only for $q\neq0$. The neutral
T-duality-inspired shadow must instead be calculated directly from the
background metric using Eq.~\eqref{eq:map_geometric_shadow}.

\section{Frequency-dependent shadow and thin-disk emission in a plasma}
\label{sec:plasma_disk}

We now examine the propagation of electromagnetic radiation through an
external plasma surrounding the T-duality-inspired regular black hole.
We follow the relativistic dispersive-transport construction of
~\cite{Perlick:2015vta,Kichenassamy:1985zz,
Lindquist:1966igj,Kobialko:2025plasma}. The plasma is assumed to be
cold, nonmagnetized, pressureless, transparent, and sufficiently dilute
that its gravitational backreaction can be neglected.

An important distinction is required in the present NED spacetime.
Test Maxwell radiation propagating through an external plasma is
described unambiguously by the background metric $g_{\mu\nu}$. By
contrast, the extraordinary electromagnetic perturbations of the NED
sector propagate, in vacuum, on an effective optical conformal class.
Adding a plasma-frequency term to the corresponding Hamiltonian makes
the result dependent on the conformal representative. A unique
extraordinary NED-plus-plasma dispersion relation therefore requires a
microscopic constitutive model combining the NED vacuum polarization
with the material response of the plasma.

We consequently organize the analysis into two parts. The first is a
test-radiation calculation based on the background metric;
it is used for the thin-disk images and spectra. The second introduces a
minimal phenomenological continuation of the extraordinary NED cone and
is used only to display how the NED optical functions could modify the
frequency-dependent shadow. Results from the second prescription should
not be interpreted as a unique material-plasma prediction.

\subsection{Cold-plasma Hamiltonian and propagation equations}
\label{subsec:plasma_hamiltonian}

The plasma frequency measured in the local rest frame of the material is
\begin{equation}
 \omega_p^2(r)
 =
 \frac{4\pi e_{\rm el}^{\,2}}{m_e}N_e(r),
 \label{eq:plasmafrequency}
\end{equation}
where $e_{\rm el}$ and $m_e$ are the electron charge and mass, and
$N_e$ is the proper electron number density. In the geometric-optics
limit, test Maxwell radiation is governed by
\begin{equation}
 {\mathscr H}_{\rm pl}
 =
 \frac{1}{2}
 \left[
 g^{\mu\nu}p_\mu p_\nu+\omega_p^2(r)
 \right]
 =0.
 \label{eq:plasmaHamiltonian}
\end{equation}
The local dispersion relation is
\begin{equation}
 k^2=\omega^2-\omega_p^2,
 \label{eq:plasmadispersion}
\end{equation}
and hence propagating rays satisfy $\omega\geq\omega_p$.

For a general static and spherically symmetric line element,
\begin{equation}
 \dd s^2
 =
 -A(r)\dd t^2+B(r)\dd r^2+C(r)\dd\Omega^2,
 \label{eq:plasmageneralmetric}
\end{equation}
the conserved energy and angular momentum in the equatorial plane are
\begin{equation}
 p_t=-\omega_\infty,
 \qquad
 p_\phi=L,
 \qquad
 b=\frac{L}{\omega_\infty}.
 \label{eq:plasmaconserved}
\end{equation}
Introducing
\begin{equation}
 p=\frac{p_r}{B\omega_\infty},
\end{equation}
and rescaling the affine parameter by $\omega_\infty$, the radial
equations become
\begin{align}
 \dot r&=p,
 \label{eq:plasmardot}\\
 \dot p&=\frac{1}{2}\frac{\dd V}{\dd r},
 \label{eq:plasmapdot}\\
 \dot\phi&=\frac{b}{C(r)},
 \qquad
 \dot t=\frac{1}{A(r)},
 \label{eq:plasmaangular}\\
 p^2&=V(r),
 \label{eq:plasmarconstraint}
\end{align}
where
\begin{equation}
 V(r)
 =
 \frac{1}{B(r)}
 \left[
 \frac{1}{A(r)}
 -\frac{b^2}{C(r)}
 -\frac{\omega_p^2(r)}{\omega_\infty^2}
 \right].
 \label{eq:plasmaeffectivepotential}
\end{equation}

For the background T-duality-inspired metric,
\begin{equation}
 A(r)=f(r),
 \qquad
 B(r)=\frac{1}{f(r)},
 \qquad
 C(r)=r^2,
\end{equation}
and therefore
\begin{equation}
 V_g(r)
 =
 1-\frac{f(r)b^2}{r^2}
 -\frac{f(r)\omega_p^2(r)}{\omega_\infty^2}.
 \label{eq:plasmaVbackground}
\end{equation}
Equations~\eqref{eq:plasmardot}--\eqref{eq:plasmaVbackground}
provide the system used below for backward ray tracing.

\subsection{Circular plasma rays and the frequency-dependent shadow}
\label{subsec:plasma_shadow_general}

Circular plasma rays satisfy
\begin{equation}
 V(r_c)=0,
 \qquad
 V'(r_c)=0.
 \label{eq:plasmacircularconditions}
\end{equation}
For the general metric \eqref{eq:plasmageneralmetric}, these conditions
give
\begin{equation}
 \omega_\infty^2
 =
 A^2
 \frac{\left(C\omega_p^2\right)'}
 {AC'-CA'},
 \label{eq:plasmacircularfrequency}
\end{equation}
and
\begin{equation}
 b_c^2
 =
 \frac{C^2}{A^2}
 \frac{\left(A\omega_p^2\right)'}
 {\left(C\omega_p^2\right)'}.
 \label{eq:plasmacircularimpact}
\end{equation}
A prime denotes differentiation with respect to $r$.

For an asymptotically flat spacetime and a plasma profile that vanishes
at infinity, the shadow radius observed at infinity is
\begin{equation}
 R_{\rm sh}^2
 =
 \left.
 \frac{C(r)}{A(r)}
 \left[
 1-
 \frac{A(r)\omega_p^2(r)}{\omega_\infty^2}
 \right]
 \right|_{r=r_c},
 \label{eq:plasmashadowgeneral}
\end{equation}
where $\omega_\infty$ is fixed by Eq.~\eqref{eq:plasmacircularfrequency}.
It equals the observed frequency only at infinity.  At a finite static
observer radius $r_o$, one instead has
$\omega_\infty=\sqrt{A(r_o)}\,\omega_o$.

The shadow disappears when the critical impact parameter vanishes.
This occurs at a radial equilibrium point $r_{\rm e}$ satisfying
\begin{equation}
 \left[A(r)\omega_p^2(r)\right]'_{r=r_{\rm e}}=0.
 \label{eq:plasmaequilibriumcondition}
\end{equation}
The corresponding asymptotic cutoff frequency is
\begin{equation}
 \omega_{\rm e}^2
 =
 A(r_{\rm e})\omega_p^2(r_{\rm e}).
 \label{eq:plasmaequilibriumfrequency}
\end{equation}
Below this frequency, rays that would otherwise probe the black hole are
reflected by the plasma before entering the near-horizon region. The
absence of a central capture shadow does not imply that the accretion
disk becomes invisible, because direct and reflected primary disk rays
may still reach the observer.

\subsection{Exact background-plasma shadow for a power-law profile}
\label{subsec:plasma_shadow_td}

We adopt the inverse power-law profile
\begin{equation}
 \omega_p^2(r)
 =
 \omega_0^2
 \left(\frac{M}{r}\right)^\sigma,
 \qquad
 \sigma>0,
 \label{eq:plasmapowerlaw}
\end{equation}
and distinguish the conserved (asymptotic) frequency from the frequency
measured by a finite static observer:
\begin{equation}
 \nu_\infty=\frac{\omega_\infty}{\omega_0},
 \qquad
 \nu_o=\frac{\omega_o}{\omega_0}
 =\frac{\nu_\infty}{\sqrt{A(r_o)}}.
 \label{eq:nudefinition}
\end{equation}
We use the variables and metric function in
Eqs.~\eqref{eq:dimensionless_variables} and \eqref{eq:fdimless}.

For $\sigma\neq2$, Eqs.~\eqref{eq:plasmacircularfrequency} and
\eqref{eq:plasmashadowgeneral} give the parametric shadow relations
\begin{equation}
 \nu_\infty^2(x)
 =
 \frac{(2-\sigma)f(x)^2}
 {2f(x)-xf_x(x)}
 \left(\frac{\mu}{x}\right)^\sigma
 \label{eq:plasmafrequencyTD}
\end{equation}
and
\begin{equation}
 \left(\frac{R_{\rm sh}^{(g)}}{\ell}\right)^2
 =
 \frac{x^2\left[xf_x(x)-\sigma f(x)\right]}
 {(2-\sigma)f(x)^2}.
 \label{eq:plasmashadowTD}
\end{equation}
The superscript $(g)$ indicates that these rays propagate according to
the background geometry.

For numerical work it is useful to define
\begin{equation}
 \Delta(x)
 =
 2f-xf_x
 =
 2-\frac{6\mu x^4}{(1+x^2)^{5/2}}
 +\frac{4e^2x^4}{(1+x^2)^3},
 \label{eq:plasmaDelta}
\end{equation}
and
\begin{equation}
\begin{aligned}
 {\cal N}_\sigma(x)
 &=
 xf_x-\sigma f
 \\
 &=
 -\sigma
 +
 \frac{
 2\mu x^2\left[(\sigma-2)+(\sigma+1)x^2\right]
 }{
 (1+x^2)^{5/2}
 }
 \\
 &\quad
 +
 \frac{
 e^2x^2\left[(2-\sigma)-(\sigma+2)x^2\right]
 }{
 (1+x^2)^3
 }.
 \label{eq:plasmaNsigma}
\end{aligned}
\end{equation}
Then
\begin{equation}
 \nu_\infty^2
 =
 \frac{(2-\sigma)f^2}{\Delta}
 \left(\frac{\mu}{x}\right)^\sigma,
 \qquad
 \left(\frac{R_{\rm sh}^{(g)}}{\ell}\right)^2
 =
 \frac{x^2{\cal N}_\sigma}
 {(2-\sigma)f^2}.
 \label{eq:plasmaCompactShadow}
\end{equation}
Only branches satisfying
\begin{equation}
 x>x_+,
 \qquad
 \nu_\infty^2>0,
 \qquad
 \left(R_{\rm sh}^{(g)}\right)^2\geq0
\end{equation}
are physically relevant.

The equilibrium radius is the outer solution of
\begin{equation}
 {\cal N}_\sigma(x_{\rm e})=0,
 \label{eq:plasmaequilibriumTD}
\end{equation}
and the cutoff frequency is
\begin{equation}
 \nu_{\infty,{\rm e}}^2
 =
 f(x_{\rm e})
 \left(\frac{\mu}{x_{\rm e}}\right)^\sigma.
 \label{eq:plasmaCutoffTD}
\end{equation}

The profile $\sigma=2$ is exceptional because
$r^2\omega_p^2$ is constant. The circular orbit therefore remains at the
background-geometric radius determined by
Eq.~\eqref{eq:map_geometric_shadow}, while the shadow has the exact
frequency dependence
\begin{equation}
 \left(\frac{R_{\rm sh}^{(g)}}{M}\right)^2
 =
 \left(\frac{R_{\rm sh}^{\rm geo}}{M}\right)^2
 -\frac{1}{\nu_\infty^2}.
 \label{eq:plasmaSigma2Shadow}
\end{equation}
The shadow therefore vanishes at
\begin{equation}
 \nu_{\infty,{\rm e}}^{(g)}
 =
 \frac{M}{R_{\rm sh}^{\rm geo}}.
 \label{eq:plasmaSigma2Cutoff}
\end{equation}
Equation~\eqref{eq:plasmaSigma2Shadow} generalizes the corresponding
Schwarzschild relation to the two-scale regular geometry.

\subsection{Minimal extraordinary NED-plasma prescription}
\label{subsec:NED_plasma_minimal}

For completeness, consider a minimal phenomenological continuation of
the extraordinary NED characteristic cone. We choose the explicit
representative \eqref{eq:map_effective_metric}, with
${\cal H}=H=\Lned_F$ and ${\cal P}=P=\Phi$.  Exterior positivity follows
analytically from Eq.~\eqref{eq:map_optical_positivity_theorem} for every
charged black hole in the family.

Appending $\omega_p^2$ to the Hamiltonian defined by this representative
gives
\begin{equation}
 {\mathscr H}_{\rm ex,pl}
 =
 \frac{1}{2}
 \left[
 g_{\rm ex}^{\mu\nu}p_\mu p_\nu+\omega_p^2
 \right].
 \label{eq:minimalNEDHamiltonian}
\end{equation}
This equation is a model assumption rather than a unique consequence of
the original NED action.

For the profile \eqref{eq:plasmapowerlaw}, the extraordinary circular
orbit is parametrized by
\begin{equation}
 \nu_\infty^2(x)
 =
 \frac{f(x)}{{\cal H}(x)}
 \left(\frac{\mu}{x}\right)^\sigma
 \frac{
 (2-\sigma)/x-{\cal P}_x/{\cal P}
 }{
 2/x-{\cal P}_x/{\cal P}
 -f_x/f+{\cal H}_x/{\cal H}
 }.
 \label{eq:minimalNEDfrequency}
\end{equation}
The corresponding shadow radius is
\begin{equation}
 \left(
 \frac{R_{\rm sh}^{\rm ex}}{\ell}
 \right)^2
 =
 \frac{x^2{\cal H}(x)}
 {f(x){\cal P}(x)}
 \left[
 1-
 \frac{f(x)}{{\cal H}(x)}
 \frac{(\mu/x)^\sigma}{\nu_\infty^2}
 \right].
 \label{eq:minimalNEDshadow}
\end{equation}
The extraordinary radial-equilibrium condition is
\begin{equation}
 \frac{f_x}{f}
 -
 \frac{{\cal H}_x}{{\cal H}}
 -
 \frac{\sigma}{x}
 =0,
 \label{eq:minimalNEDequilibrium}
\end{equation}
with
\begin{equation}
 \nu_{\infty,{\rm e}}^{2}
 =
 \frac{f(x_{\rm e})}{{\cal H}(x_{\rm e})}
 \left(\frac{\mu}{x_{\rm e}}\right)^\sigma.
 \label{eq:minimalNEDcutoff}
\end{equation}

The high-frequency behavior can be written compactly.  The vacuum
function $Q(x)$ is precisely ${\cal B}(x)$ from
Eq.~\eqref{eq:map_impact_function}; define only
\begin{equation}
 D_\sigma(x)
 =
 \frac{\mu^\sigma x^{2-\sigma}}{{\cal P}(x)}.
 \label{eq:minimalNEDD}
\end{equation}
Let $x_0$ be the vacuum NED light-ring radius, so that
${\cal B}'(x_0)=0$. Then
\begin{equation}
 x_c
 =
 x_0+
 \frac{D_\sigma'(x_0)}
 {\nu_\infty^2{\cal B}''(x_0)}
 +{\cal O}(\nu_\infty^{-4}),
 \label{eq:minimalNEDorbitExpansion}
\end{equation}
and
\begin{equation}
 \left(
 \frac{R_{\rm sh}^{\rm ex}}{\ell}
 \right)^2
 =
 \left(
 \frac{R_{\rm sh}^{\rm NED}}{\ell}
 \right)^2
 -
 \frac{
 \mu^\sigma x_0^{2-\sigma}
 }{
 {\cal P}(x_0)\nu_\infty^2
 }
 +{\cal O}(\nu_\infty^{-4}).
 \label{eq:minimalNEDshadowExpansion}
\end{equation}
The leading displacement of the orbit does not contribute explicitly to
the leading shadow correction because ${\cal B}$ is stationary at $x_0$.

\subsection{Circular disk motion and Novikov--Thorne flux}
\label{subsec:plasma_disk_flux}

The accretion flow is treated as test matter following timelike
geodesics of the background spacetime. The disk is geometrically thin,
optically thick, and located in the equatorial plane. Its angular
velocity, specific energy, and specific angular momentum are
\cite{Page:1974he,Bambi:2017khi}
\begin{align}
 \Omega_D^2(r)
 &=
 \frac{f'(r)}{2r},
 \label{eq:diskOmega}\\
 E_D(r)
 &=
 \frac{\sqrt{2}\,f(r)}
 {\sqrt{2f(r)-rf'(r)}},
 \label{eq:diskEnergy}\\
 L_D(r)
 &=
 \sqrt{
 \frac{r^3f'(r)}
 {2f(r)-rf'(r)}
 }.
 \label{eq:diskMomentum}
\end{align}
The disk four-velocity is
\begin{equation}
 u_D^\mu
 =
 \frac{
 \delta_t^\mu+\Omega_D\delta_\phi^\mu
 }{
 \sqrt{f-r^2\Omega_D^2}
 }.
 \label{eq:diskvelocity}
\end{equation}

The innermost stable circular orbit is the outer solution of
\begin{equation}
 3ff'
 +rff''
 -2r(f')^2=0.
 \label{eq:TDISCO}
\end{equation}
The Novikov--Thorne energy flux from one surface of the disk is
\begin{equation}
\begin{aligned}
 {\cal F}_D(r)
 ={}&
 -\frac{\dot M}{4\pi r}
 \frac{\Omega_D'(r)}
 {\left[E_D(r)-\Omega_D(r)L_D(r)\right]^2}
 \\
 &\times
 \int_{r_{\rm ISCO}}^r
 \left[E_D-\Omega_DL_D\right]
 L_D'(\tilde r)\,\dd\tilde r.
 \label{eq:TDdiskflux}
\end{aligned}
\end{equation}
Here $\sqrt{-\det g_{3D}}=r$ for the $(t,r,\phi)$ submetric.
The effective surface temperature is
\begin{equation}
 T_D(r)
 =
 \left[
 \frac{{\cal F}_D(r)}
 {\sigma_{\rm SB}}
 \right]^{1/4},
 \qquad
 \sigma_{\rm SB}=\frac{\pi^2}{60}.
 \label{eq:TDdisktemperature}
\end{equation}
The nominal radiative efficiency is
\begin{equation}
 \eta_{\rm NT}=1-E_D(r_{\rm ISCO}).
 \label{eq:TDefficiency}
\end{equation}

\subsection{Redshift, intensity transport, and observed spectrum}
\label{subsec:plasma_transport}

We assume that the plasma co-rotates with the disk near the emitting
surface, $\Omega_p=\Omega_D$, and becomes static at the observer.  For an
inclined ray, the Doppler factor depends on the axial impact parameter
$\lambda_z=L_z/\omega_\infty$, not on the total impact parameter $b$ that
enters the radial potential.  The redshift factor is
\begin{equation}
 1+z
 =
 \left(1-\Omega_D\lambda_z\right)
 \sqrt{
 \frac{f(r_o)}
 {f(r)-r^2\Omega_D^2}
 }.
 \label{eq:TDplasmaredshift}
\end{equation}

In a transparent medium with negligible absorption and no intrinsic
plasma emission, the invariant distribution function is conserved along
the ray. The observed specific intensity is therefore
\cite{Lindquist:1966igj,Kichenassamy:1985zz,
Kobialko:2025plasma}
\begin{equation}
 I_o(\omega_o)
 =
 \frac{\omega_o^3}{4\pi^3}
 \frac{
 1-\bar\omega_p^2/\omega_o^2
 }{
 \exp\left[
 (1+z)\omega_o/T_D
 \right]-1
 }.
 \label{eq:TDplasmaIntensity}
\end{equation}
The bar denotes evaluation at the observer; $\bar\omega_p\to0$ for an
asymptotic observer when the plasma profile vanishes at infinity.

The observed spectral flux is
\begin{equation}
 F_o(\omega_o)
 =
 \int_{\rm screen}
 I_o(\omega_o)\,\dd\Omega.
 \label{eq:TDobservedflux}
\end{equation}
Using stereographic screen coordinates,
\begin{equation}
 X=2\tan\left(\frac{\Theta}{2}\right)\cos\Psi,
 \qquad
 Y=2\tan\left(\frac{\Theta}{2}\right)\sin\Psi,
\end{equation}
the solid-angle element becomes
\begin{equation}
 \dd\Omega
 =
 \frac{16\,\dd X\,\dd Y}
 {(4+X^2+Y^2)^2}.
 \label{eq:TDsolidangle}
\end{equation}

\subsection{Numerical implementation and representative results}
\label{subsec:plasma_numerical_results}

To illustrate the combined influence of the regular geometry and the
external plasma, we consider the representative parameters
\begin{equation}
 \frac{\ell}{M}=0.2,
 \qquad
 \frac{q}{M}=0.3.
 \label{eq:plasmaBenchmark}
\end{equation}
These ratios imply $(\mu,e)=(5,1.5)$ and satisfy
$3\mu=15>2e^2=4.5$.  Exterior optical positivity then follows from the
analytic theorem \eqref{eq:map_optical_positivity_theorem}, rather than
from a pointwise numerical scan.

For this configuration, the principal radii are
\begin{align}
 \frac{r_+}{M}
 &=
 1.92213,
 \label{eq:benchmarkHorizon}\\
 \frac{r_{\rm ISCO}}{M}
 &=
 5.79657,
 \label{eq:benchmarkISCO}\\
 \frac{r_{\rm ph}^{\rm geo}}{M}
 &=
 2.90359,
 \qquad
 \frac{R_{\rm sh}^{\rm geo}}{M}
 =
 5.08039,
 \label{eq:benchmarkGeoShadow}\\
 \frac{r_{\rm ph}^{\rm NED}}{M}
 &=
 2.87304,
 \qquad
 \frac{R_{\rm sh}^{\rm NED}}{M}
 =
 4.55835.
 \label{eq:benchmarkNEDShadow}
\end{align}
Thus, for this benchmark, the nonlinear-electrodynamic optical response
reduces the vacuum shadow radius by approximately
\begin{equation}
 \frac{
 R_{\rm sh}^{\rm geo}-R_{\rm sh}^{\rm NED}
 }{
 R_{\rm sh}^{\rm geo}
 }
 \simeq10.3\%.
 \label{eq:benchmarkShadowReduction}
\end{equation}

The disk energy at the ISCO and the corresponding efficiency are
\begin{equation}
 E_{\rm ISCO}=0.941011,
 \qquad
 \eta_{\rm NT}=0.058989.
 \label{eq:benchmarkEfficiency}
\end{equation}
Table~\ref{tab:disk_benchmark} compares the disk with Schwarzschild and
Reissner--Nordstr\"om black holes at the same $M$ and $q$.

\begin{table*}[t]
 \caption{
 Circular-orbit and Novikov--Thorne quantities for the representative
 model \eqref{eq:plasmaBenchmark}. The flux is quoted in geometrized
 units and normalized by the accretion rate.
 }
 \label{tab:disk_benchmark}
 \begin{ruledtabular}
 \begin{tabular}{lcccc}
 Geometry
 &
 $r_{\rm ISCO}/M$
 &
 $\eta_{\rm NT}$
 &
 $r_{{\cal F},\max}/M$
 &
 $M^2{\cal F}_{\max}/\dot M$
 \\
 \hline
 Schwarzschild
 &
 $6.00000$
 &
 $0.057191$
 &
 $9.55103$
 &
 $1.36781\times10^{-5}$
 \\
 Reissner--Nordstr\"om
 &
 $5.86278$
 &
 $0.058413$
 &
 $9.33354$
 &
 $1.46204\times10^{-5}$
 \\
 T-duality NED
 &
 $5.79657$
 &
 $0.058989$
 &
 $9.23075$
 &
 $1.50903\times10^{-5}$
 \end{tabular}
 \end{ruledtabular}
\end{table*}

Relative to Schwarzschild, the zero-point-length and charge corrections
move the ISCO inward by approximately $3.39\%$, increase the nominal
efficiency by $3.14\%$, and increase the maximum local disk flux by
approximately $10.3\%$.

The plasma ray tracing was performed for
\begin{equation}
 r_o=40M,
 \qquad
 \theta_o=84^\circ,
 \qquad
 \dot M=0.1,
 \label{eq:raytraceObserver}
\end{equation}
with the disk extending from $r_{\rm ISCO}$ to $20M$ and a background
termination sphere at $50M$. The full three-dimensional Hamilton system
given in Appendix~\ref{app:raytracing} was integrated backward using a
fourth-order Runge--Kutta method; its radial reduction agrees with
Eqs.~\eqref{eq:plasmardot}--\eqref{eq:plasmaVbackground}. The plasma was
taken to be transparent and co-rotating with the disk near the emitting
surface.

For the $\sigma=2$ profile, Eq.~\eqref{eq:plasmaSigma2Cutoff} gives
\begin{equation}
 \nu_{\infty,{\rm e}}^{(g)}
 =
 0.19684.
 \label{eq:benchmarkCutoff2}
\end{equation}
The minimal extraordinary prescription instead gives
\begin{equation}
 \nu_{\infty,{\rm e}}^{\rm ex}
 \simeq0.13778.
 \label{eq:benchmarkNEDCutoff2}
\end{equation}
For $\sigma=6$, the corresponding values are
\begin{equation}
 \nu_{\infty,{\rm e}}^{(g)}
 \simeq0.03274,
 \qquad
 \nu_{\infty,{\rm e}}^{\rm ex}
 \simeq0.02128.
 \label{eq:benchmarkCutoff6}
\end{equation}
The lower cutoff for the rapidly decreasing profile reflects the strong
localization of the plasma response near the black hole.  For the
background branch at $r_o=40M$, $f(r_o)=0.950058$; the corresponding
locally measured cutoff frequencies are
$\nu_{o,{\rm e}}^{(g)}=0.20194$ for $\sigma=2$ and $0.03359$ for
$\sigma=6$.

\begin{figure}[t]
 \centering
 \includegraphics[width=\columnwidth]{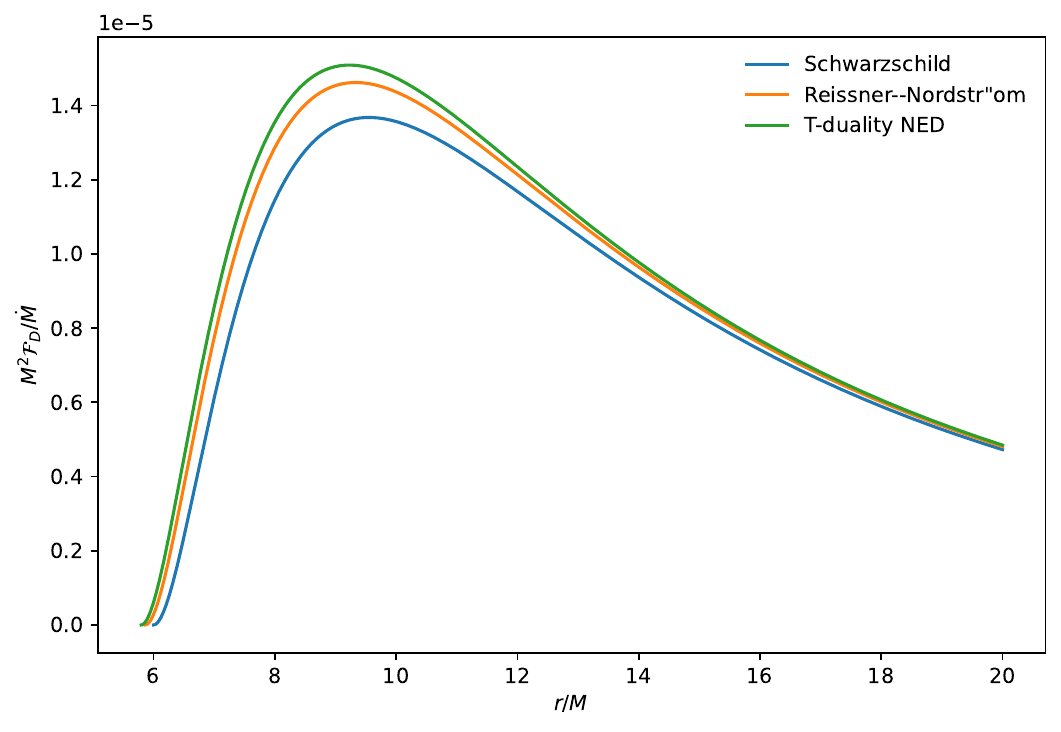}
 \caption{
 Novikov--Thorne energy flux for Schwarzschild,
 Reissner--Nordstr\"om, and the T-duality-inspired NED geometry with
 $\ell/M=0.2$ and $q/M=0.3$. The curves are normalized by $\dot M$.
 }
 \label{fig:TDdiskFlux}
\end{figure}

Figure~\ref{fig:TDdiskFlux} shows that the T-duality-inspired disk is
slightly hotter and more centrally concentrated than the corresponding
Schwarzschild and RN disks.

\begin{figure}[t]
 \centering
 \includegraphics[width=\columnwidth]{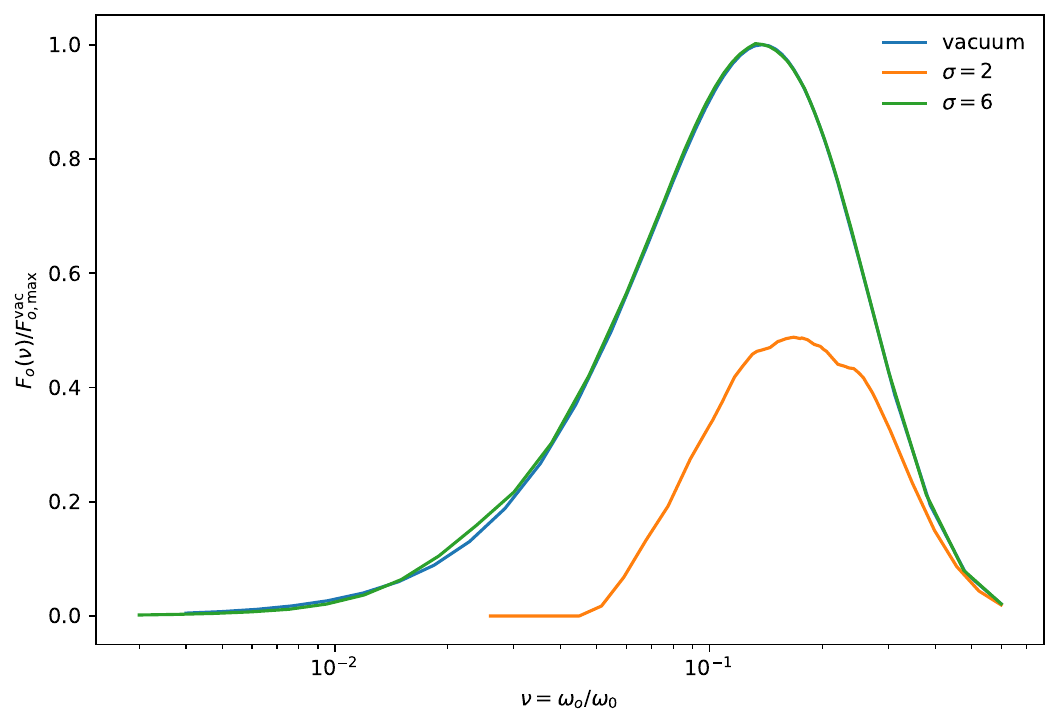}
 \caption{
 Spectral flux observed at $r_o=40M$ and
 $\theta_o=84^\circ$ for the benchmark
 $\ell/M=0.2$, $q/M=0.3$, and $\dot M=0.1$.
 The horizontal coordinate is the conserved ratio
 $\nu_\infty=\omega_\infty/\omega_0$, and the
 fluxes are divided by the maximum vacuum flux. The shallow
 $\sigma=2$ distribution produces a substantial cutoff, peak shift,
 and flux suppression, whereas the $\sigma=6$ spectrum remains close
 to its vacuum counterpart.
 }
 \label{fig:TDplasmaSpectrum}
\end{figure}

The frequency-dependent spectral flux is shown in
Fig.~\ref{fig:TDplasmaSpectrum}.  The illustrative normalization uses
$M\omega_0=1$.  Table~\ref{tab:plasma_spectral_peaks} reports the
conserved frequency coordinate and normalizes every peak to the vacuum
maximum.
\begin{table}[t]
 \caption{Peak locations and relative heights for the illustrative
 spectrum with $M\omega_0=1$.}
 \label{tab:plasma_spectral_peaks}
 \begin{ruledtabular}
 \begin{tabular}{lcc}
 Model & $\nu_{\infty,{\rm peak}}$ &
 $F_{o,\max}/F_{o,\max}^{\rm vac}$ \\
 \hline
 Vacuum & $0.138$ & $1$ \\
 $\sigma=2$ & $0.168$ & $0.488$ \\
 $\sigma=6$ & $0.136$ & $1.00$ \\
 \end{tabular}
 \end{ruledtabular}
\end{table}
For $\sigma=2$, the gradual plasma profile shifts the spectral maximum upward
by approximately $21\%$ while suppressing its amplitude by approximately
$51\%$. This suppression occurs even though plasma absorption has been
neglected; it is caused by dispersive reflection and the reduced screen
region connected to the disk.

For $\sigma=6$, the spectrum differs from the vacuum result by less than one percent
near its maximum. A steep density profile affects only a comparatively
small strong-field region and consequently produces a much weaker
integrated spectral signature.

\begin{figure*}[t]
 \centering
 \includegraphics[width=0.92\textwidth]{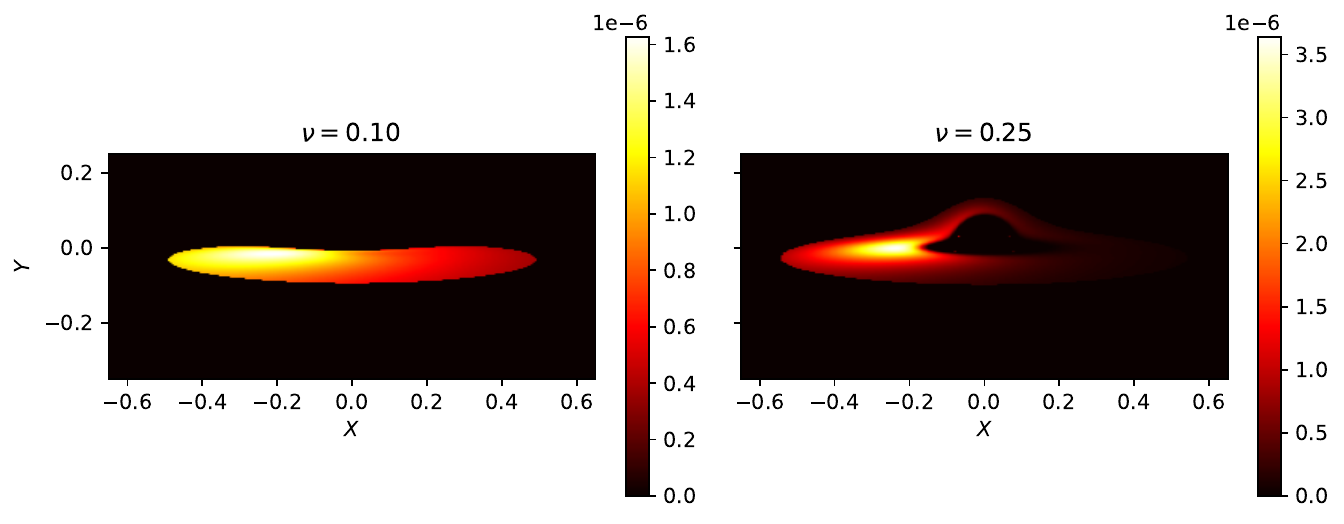}
 \caption{
 Specific-intensity maps for the power-law plasma profile
 $\omega_p^2=\omega_0^2(M/r)^2$, with
 $\ell/M=0.2$, $q/M=0.3$, $r_o=40M$,
 $\theta_o=84^\circ$, and $\dot M=0.1$.
 Left: $\nu_\infty=0.10$, below the shadow-disappearance frequency.
 Right: $\nu_\infty=0.25$, above the cutoff. Each color scale gives the
 intensity in the numerical geometrized normalization and is shown
 separately because the two frequency slices have different maxima.
 }
 \label{fig:TDplasmaTransition}
\end{figure*}

Figure~\ref{fig:TDplasmaTransition} displays the image transition for
$\sigma=2$. At
$\nu_\infty=0.10<\nu_{\infty,{\rm e}}^{(g)}$, the central capture shadow
and higher-order lensed arcs are absent because the plasma reflects rays
before they probe the photon region. The observer sees primarily a
flattened direct image of the approaching disk. At
$\nu_\infty=0.25>\nu_{\infty,{\rm e}}^{(g)}$, a central silhouette and a lensed upper
arc reappear. The strong left--right intensity asymmetry is generated by
the Doppler factor in Eq.~\eqref{eq:TDplasmaredshift}.

The numerical results exhibit two distinct plasma regimes. Slowly
decaying profiles, exemplified by $\sigma=2$, affect a broad range of
radii and produce an observable cutoff, a strong frequency dependence of
the image morphology, and a sizable suppression of the integrated
spectrum. Rapidly decaying profiles, such as $\sigma=6$, primarily
modify rays near the compact object and leave the disk-integrated
spectrum almost unchanged except at sufficiently low frequencies.

The choice $M\omega_0=1$ fixes the ratio of the plasma scale to the disk
temperature only for this illustration. Converting $\nu_\infty$ into an
observing frequency, or changing the spectral normalization, requires an
astrophysical model for the electron density, composition, mass, and
accretion rate. The benchmark should therefore not be interpreted as a
direct constraint on M87$^*$ or Sgr~A$^*$ without an independent plasma
calibration.

\section{Conclusions}
\label{sec:discussion}

We have constructed a two-scale, magnetically charged regular geometry
supported, for $q\neq0$, by an inverse-reconstructed NED source.  The
background has the neutral zero-point-length, Reissner--Nordstr\"om,
Ay\'on-Beato--Garc\'ia, and Schwarzschild limits described in
Sec.~\ref{sec:solution}.  All curvature invariants are finite for
$\ell>0$.  The sign of $2M\ell-q^2$ fixes the de Sitter, Minkowski, or
anti-de Sitter character of the core, whereas the stronger condition
\eqref{eq:globalwec} is necessary and sufficient for the WEC at every
radius.

The reconstructed $\Lned(F)$ is single valued on the magnetic branch,
approaches Maxwell theory at weak field, and remains finite at strong
field.  Its noninteger weak-field powers and explicit dependence on
$(M,q,\ell)$ are nevertheless important: varying those parameters
generally varies the effective matter theory.  The family should not be
interpreted as a continuous state space of one universal microscopic NED
without an additional construction.

The horizon analysis gives a unique extremal radius for each $|q|/\ell$.
The physical Hawking temperature vanishes there, while the Einstein--Hilbert
Wald entropy retains the area law.  Direct differentiation of the horizon
mass produces the exact state differential \eqref{eq:exactstatedifferential}
and the physical-temperature horizon identity
\eqref{eq:modifiedfirstlawfull}, with the homogeneous Smarr relation
\eqref{eq:modifiedSmarr}.  The exact heat capacity
\eqref{eq:heatcapacityexplicit} identifies the locally stable and unstable
branches.  The Penrose-type entropy inequality and the two specific-heat
bounds remain valid on the outer branch, whereas the scale-free quantity
$Z$ is negative for sufficiently large horizons when $\ell\neq0$.

A key analytic result is the exterior optical-admissibility theorem
\eqref{eq:map_optical_positivity_theorem}: every charged black hole in the
family satisfies $\Lned_F>0$ and $\Phi>0$ throughout its domain of outer
communication.  The zero of $\Phi$ found near the regular center on the
WEC branch is always hidden behind the event horizon and is not an
additional parameter-space boundary.  The extraordinary capture shadow is
therefore well defined by the global minimum in
Eq.~\eqref{eq:map_ned_shadow}.  This global prescription supersedes the
selection of an unspecified local or ``outermost'' stationary point and
separates the NED propagation effect from the background-geodesic shadow.
The fixed-$q\neq0$ expansion about RN quantifies this separation in
Eqs.~\eqref{eq:NEDshadowperturbative} and
\eqref{eq:NEDminusgeometricshadow}.

The weak-field sector supplies complementary tests.  Neutral massive
bodies follow background timelike geodesics, giving the periapsis advance
\eqref{eq:ct_periapsis_weak}.  Background light bending through fourth
post-Minkowskian order is given by Eq.~\eqref{eq:ct_bending_weak}, while
the finite-endpoint time delay and redshift are
Eqs.~\eqref{eq:ct_delay_finite} and \eqref{eq:ct_redshift_weak}.  The
zero-point length does not shift the standard first-post-Newtonian
parameter $\gamma$; it first enters the isotropic metric at order
$\rho^{-3}$.  Consequently, the Solar-System values in
Table~\ref{tab:ct_sensitivities} are conditional sensitivities to exterior
metric coefficients, not fitted constraints on a magnetically charged
solar model.  Extraordinary NED rays require the separate characteristic
integrals \eqref{eq:ct_optical_bending} and
\eqref{eq:ct_optical_delay}, whose $q\to0$ limit is nonuniform.

For an external cold plasma, minimally coupled test radiation has an
unambiguous background Hamiltonian.  The power-law model admits the
parametric shadow relations \eqref{eq:plasmaCompactShadow}; for
$\sigma=2$, Eq.~\eqref{eq:plasmaSigma2Shadow} is exact.  The benchmark
reproduces the horizon, photon, ISCO, efficiency, and disk-flux values in
Eqs.~\eqref{eq:benchmarkHorizon}--\eqref{eq:benchmarkEfficiency} and
Table~\ref{tab:disk_benchmark}.  At a finite observer, the locally measured
frequency must be distinguished from the conserved ratio $\nu_\infty$ as
in Eq.~\eqref{eq:nudefinition}; the three-dimensional initialization in
Appendix~\ref{app:raytracing} likewise distinguishes the total impact
parameter from $L_z/E$ in the Doppler factor.  The plotted spectra use the
explicit illustrative normalization $M\omega_0=1$.  Translating them to
M87$^*$ or Sgr~A$^*$ requires an electron-density normalization, a
composition model, and an astrophysical accretion prescription.

The extraordinary NED--plasma Hamiltonian in
Sec.~\ref{subsec:NED_plasma_minimal} remains deliberately phenomenological.
A material plasma frequency breaks the conformal freedom of the vacuum NED
characteristic metric; a unique prediction therefore requires a
microscopic constitutive theory combining the nonlinear vacuum response
with the material dielectric response.  Natural extensions are the
electric Hamiltonian formulation, perturbative stability of the coupled
gravitational--electromagnetic system, calibrated radiative-transfer fits,
and rotating solutions obtained without assuming that a
Newman--Janis-type construction preserves spacetime or optical regularity.

\acknowledgments
A. \"O and R. P. would like to acknowledge the networking support of the COST Action CA21106 - COSMIC WISPers in the Dark Universe: Theory, astrophysics and experiments (CosmicWISPers), the COST Action CA22113 - Fundamental challenges in theoretical physics (THEORY-CHALLENGES), the COST Action CA21136 - Addressing observational tensions in cosmology with systematics and fundamental physics (CosmoVerse), the COST Action CA23130 - Bridging high and low energies in search of quantum gravity (BridgeQG), and the COST Action CA23115 - Relativistic Quantum Information (RQI) funded by COST (European Cooperation in Science and Technology). J. S. acknowledges the FONDECYT
grant N°1220065, Chile.

\appendix

\section{Three-dimensional plasma ray tracing}
\label{app:raytracing}

For an inclined observer, the equatorial radial reduction is insufficient
to compute disk intersections and Doppler shifts.  In background spherical
coordinates, the full Hamiltonian \eqref{eq:plasmaHamiltonian} reads
\begin{equation}
 2{\mathscr H}_{\rm pl}
 =-\frac{E^2}{f}+fp_r^2
 +\frac{p_\theta^2}{r^2}
 +\frac{L_z^2}{r^2\sin^2\theta}
 +\omega_p^2=0.
 \label{eq:app_full_hamiltonian}
\end{equation}
Besides $E=-p_t$ and $L_z=p_\phi$, spherical symmetry gives the conserved
total angular momentum
\begin{equation}
 K=p_\theta^2+\frac{L_z^2}{\sin^2\theta},
 \qquad
 b^2=\frac{K}{E^2},
 \qquad
 \lambda_z=\frac{L_z}{E}.
 \label{eq:app_angular_constants}
\end{equation}
Hamilton's equations required for backward ray tracing are
\begin{align}
 \dot r&=fp_r,
 &\dot\theta&=\frac{p_\theta}{r^2},
 \nonumber\\
 \dot\phi&=\frac{L_z}{r^2\sin^2\theta},
 \label{eq:app_coordinate_equations}\\
 \dot p_r&=-\frac{1}{2}\left[
 \frac{E^2f'}{f^2}+f'p_r^2-\frac{2K}{r^3}
 +(\omega_p^2)'
 \right],
 \label{eq:app_radial_momentum}\\
 \dot p_\theta&=
 \frac{L_z^2\cos\theta}{r^2\sin^3\theta}.
 \label{eq:app_polar_momentum}
\end{align}
Their radial first integral is
$\dot r^{\,2}=E^2-f(K/r^2+\omega_p^2)$, which reduces to
Eq.~\eqref{eq:plasmaVbackground} after division by $E^2$.

At a static observer, use the background orthonormal tetrad
\begin{equation}
 \begin{aligned}
 e_{\hat t}&=f_o^{-1/2}\partial_t,
 &e_{\hat r}&=f_o^{1/2}\partial_r,
 \\
 e_{\hat\theta}&=r_o^{-1}\partial_\theta,
 &e_{\hat\phi}&=(r_o\sin\theta_o)^{-1}\partial_\phi.
 \end{aligned}
 \label{eq:app_observer_tetrad}
\end{equation}
Let $k_o=\sqrt{\omega_o^2-\bar\omega_p^2}=n_o\omega_o$ and orient
$(\Theta,\Psi)$ from the inward radial direction.  A convenient screen
initialization is
\begin{equation}
 \begin{aligned}
 p^{\hat r}&=-k_o\cos\Theta,
 \\
 p^{\hat\theta}&=k_o\sin\Theta\cos\Psi,
 \\
 p^{\hat\phi}&=k_o\sin\Theta\sin\Psi.
 \end{aligned}
 \label{eq:app_local_momentum}
\end{equation}
It implies
\begin{align}
 E&=\sqrt{f_o}\,\omega_o,
 &p_r&=\frac{p^{\hat r}}{\sqrt{f_o}},
 \nonumber\\
 p_\theta&=r_o\,p^{\hat\theta},
 &L_z&=r_o\sin\theta_o\,p^{\hat\phi},
 \label{eq:app_initial_covectors}\\
 b^2&=\frac{r_o^2n_o^2\sin^2\Theta}{f_o},
 \nonumber\\
 \lambda_z&=
 \frac{r_o\sin\theta_o\,n_o\sin\Theta\sin\Psi}{\sqrt{f_o}}.
 \label{eq:app_screen_constants}
\end{align}
Disk emission is recorded at crossings of $\theta=\pi/2$ within the
chosen radial disk interval.  Equation~\eqref{eq:TDplasmaredshift} then
uses $\lambda_z$, whereas the radial capture/reflection problem uses $b$.
This distinction is essential away from an equatorial observer.

\end{document}